\documentclass[twocolumn, times, trackchanges]{aastex63_bren}

\usepackage[ruled,vlined]{algorithm2e}
\usepackage{graphicx}
\usepackage{bbm}
\usepackage{float}
\usepackage{xspace}
\usepackage{listings}
\usepackage[hypertexnames=false]{hyperref}
\usepackage{subfiles} 
\usepackage[encapsulated]{CJK}

\hypersetup{linkcolor=red,citecolor=green,filecolor=cyan,urlcolor=magenta}

\defcitealias{baron19}{B19}
\defcitealias{baron24}{B24}

\newcommand{\halpha}{H$\alpha$\xspace}
\newcommand{\hbeta}{H$\beta$\xspace}

\newcommand{\oiii}{$\text{[O\,{\sc iii}]}$\xspace}

\newcommand{\oi}{$\text{[O\,{\sc i}]}$\xspace}

\newcommand{\oiiihbeta}{$\log (\text{[O\,{\sc iii}]}/\text{H}\beta)$\xspace}
\newcommand{\niihalpha}{$\log (\text{[N\,{\sc ii}]}/\text{H}\alpha)$\xspace}
\newcommand{\siihalpha}{$\log (\text{[S\,{\sc ii}]}/\text{H}\alpha)$\xspace}
\newcommand{\oihalpha}{$\log (\text{[O\,{\sc i}]}/\text{H}\alpha)$\xspace}

\newcommand{\mic}{$\mathrm{\mu m}$\xspace}

\shorttitle{starburst to post-starburst survey}
\shortauthors{Baron et al.}

\begin{document}

\title{Sparks II: Panchromatic SED modeling and galaxy physical properties across the starburst to post-starburst sequence}

\author[0000-0003-4974-3481]{Dalya Baron}
\email{dalyabaron@gmail.com}
\affiliation{Kavli Institute for Particle Astrophysics \& Cosmology, Stanford University, CA 94305, USA}
\affiliation{Center for Decoding the Universe, Stanford University, CA 94305, USA}

\author[0000-0003-4075-7393]{David J. Setton}\thanks{Brinson Prize Fellow}
\affiliation{Department of Astrophysical Sciences, Princeton University, 4 Ivy Lane, Princeton, NJ 08544, USA}

\author[0000-0002-0463-9528]{Yilun Ma (\begin{CJK*}{UTF8}{gbsn}马逸伦\ignorespacesafterend\end{CJK*})}
\affiliation{Department of Astrophysical Sciences, Princeton University, 4 Ivy Lane, Princeton, NJ 08544, USA}

\author[0000-0002-7738-6875]{J.~X.~Prochaska}
\affiliation{Department of Astronomy and Astrophysics, University of California Santa Cruz, 1156 High Street, Santa Cruz, CA 95064, USA}
\affiliation{Kavli Institute for the Physics and Mathematics of the Universe (Kavli IPMU), 5-1-5 Kashiwanoha, Kashiwa, 277-8583, Japan}
\affiliation{Division of Science, National Astronomical Observatory of Japan, 2-21-1 Osawa, Mitaka, Tokyo 181-8588, Japan}

\author[0000-0003-4949-7217]{Ric Davies}
\affiliation{Max-Planck-Institut f\"ur extraterrestrische Physik, Giessenbachstra{\ss}e, 85748 Garching, Germany}

\author[0000-0002-5612-3427]{Jenny~E.~Greene}
\affiliation{Department of Astrophysical Sciences, Princeton University, Princeton, NJ 08544, USA}

\author[0000-0003-0291-9582]{Dieter Lutz}
\affiliation{Max-Planck-Institut f\"ur extraterrestrische Physik, Giessenbachstra{\ss}e, 85748 Garching, Germany}



\begin{abstract}

The \textit{Sparks} survey provides rest-frame near-infrared spectroscopy for 93 local massive galaxies spanning the rapid transition from starburst to post-starburst, including Balmer-strong galaxies as well as systems with active galactic nuclei (AGN). Interpreting these extreme systems requires reliable physical properties, yet these can vary substantially when derived from rest-frame optical spectroscopy versus multi-wavelength photometry, and across different fitting codes and assumptions. We assemble far-ultraviolet to far-infrared photometry for the \textit{Sparks} sample and compare the resulting galaxy properties across data types and modeling approaches, identifying the final measurements adopted for the survey. With stellar masses recovered relatively robustly, we focus on the more model-dependent quantities of star formation rates (SFRs) and histories (SFHs), and AGN activity. Fits to optical stellar continuum alone, dominated by strong Balmer absorption, systematically favor rapidly declining SFHs and suppress ongoing star formation. Benchmarking against H$\alpha$-based SFRs in the star-forming \textit{Sparks} galaxies shows that \texttt{Prospector} fits to the optical continuum spectroscopy underestimate the SFR by 0.76 dex (scatter 0.42 dex), whereas panchromatic SED-based SFRs perform better, with a $-0.15$ dex offset and 0.14 dex scatter. We therefore adopt the panchromatic SED?based SFRs for composite and AGN hosts, finding that many exhibit higher levels of star formation than previously inferred. Finally, we test the AGN torus model in \texttt{Prospector}, finding that it successfully distinguishes optically-classified AGN from star-forming galaxies, but yields torus luminosities an order of magnitude below expectations from AGN bolometric luminosities, possibly indicating intrinsically low covering factors in \textit{Sparks} AGN shaped by black-hole feedback during coalescence. 

\end{abstract}


\section{Introduction}\label{sec:intro}

Probing galaxy evolution across cosmic time requires modeling their spectra and multi-wavelength spectral energy distributions (SEDs), which encode the combined emission from stars, gas, and dust (see review by \citealt{conroy13}). Stellar population synthesis (SPS) fitting relies on ingredients such as the initial mass function, stellar isochrones, and stellar SEDs to infer stellar ages, metallicities, kinematics, and detailed star formation histories (SFHs) from absorption features and continuum shapes (e.g., \citealt{cid_fernandes05, koleva09, conroy13, wilkinson17}). Extending the wavelength range, from the ultraviolet to the radio, allows additional physical processes to be modeled alongside the stellar populations, including dust attenuation and re-emission, and in some cases, active galactic nuclei (AGN; e.g., \citealt{daCunha08, noll09, leja17, carnall18, johnson21}). This broader coverage places SPS modeling within the full energy budget of a galaxy, enforcing energy balance between ultraviolet?optical photons absorbed by dust and re-emitted in the infrared, ultimately reducing degeneracies between parameters and improving constraints on stellar masses, star formation rates (SFRs), and galaxy SFHs.

Within SED modeling frameworks, star formation histories can be modeled using either parametric or nonparametric approaches (see introduction in \citealt{leja19}). Parametric models employ functional forms such as exponentially declining, delayed-tau, or double power-law SFHs that reduce galaxy evolution to a small number of physically motivated parameters (e.g., \citealt{bell01, maraston10, papovich11, wuyts11, gladders13}). While computationally efficient, these models impose restrictive priors on allowed SFH shapes, ruling out entire classes of evolutionary scenarios. This could potentially bias or underestimate the uncertainties on derived masses and SFRs for galaxies with complex histories (e.g., \citealt{simha14, haskell24}). Nonparametric approaches discretize the SFH into time bins without assuming a functional form, enabling flexible modeling of starbursts, quenching episodes, and rejuvenation events (e.g., \citealt{kelson14, carnall18, chauke18, leja19}). However, this flexibility requires careful selection of prior probability distributions, as the choice of prior?rather than photometric noise?primarily determines uncertainties in derived properties (\citealt{leja19}).

Several popular SED fitting codes are widely used in the literature, including \texttt{MAGPHYS} \citep{daCunha08}, \texttt{BAGPIPES} \citep{carnall18}, \texttt{CIGALE} \citep{noll09, boquien19}, and \texttt{Prospector} \citep{leja17, johnson21}, which differ not only in their treatment of SFHs but also in their stellar evolutionary models, dust attenuation curves and geometry, inclusion of AGN emission, and ability to fit photometry alone versus combined spectro-photometric data. Given these different modeling choices, it is important to benchmark the different codes on diverse samples of galaxies, as well as on simulated galaxies with known properties, to identify their regimes of applicability and potential systematic biases. Several such comparison studies have been conducted in recent years (\citealt{best23, cappellari23, pacifici23, haskell24, woo24}). These studies generally find good agreement on stellar masses (\citealt{best23, pacifici23}), but significant discrepancies emerge for SFRs and dust attenuation ($\sim 0.3$ dex; dominated by galaxies on the star-forming main sequence), with codes displaying strong dust--age degeneracy (\citealt{pacifici23}). 

Galaxies along the starburst-to-post-starburst sequence occupy a distinct region of parameter space in panchromatic SED fitting. This sequence encompasses systems at different stages of galaxy quenching, from peak starburst activity, through rapid decline, to final quiescence (\citealt{wild07, peng15, maltby18, pawlik18, rowlands18, belli19, wild20, tacchella22}). Post-starburst galaxies represent a key intermediate phase in the sequence, caught rapidly quenching over hundreds of millions of years (Myr; \citealt{dressler83, couch87, poggianti99, wild07, yesuf14, alatalo16a, french18, pawlik18, baron22, french21}). 

This sequence is thus characterized by extreme properties: intense and rapidly changing SFRs, possible merger or AGN activity, and significant dust obscuration (\citealt{zabludoff96, canalizo00, yang04, yang06, kaviraj07, wild09, yesuf14, cales15, wild16, almaini17}). Depending on their stage in the transition, their optical spectra may exhibit strong Balmer absorption features from intermediate-age stellar populations, providing unique diagnostic leverage on SFHs over the past $\sim 1$ Gyr. Together, these characteristics make the starburst-to-post-starburst sequence a regime where the choice of fitting strategy is particularly consequential (\citealt{french18, wild20, suess22sfh}): inferred stellar ages, masses, and quenching timescales can differ substantially depending on the fitting code, whether spectroscopy or photometry is used, the adopted stellar libraries, and the assumed SFH priors.

In a companion paper (hereafter Paper I; Baron et al. submitted), we present \textit{Sparks}, a Magellan-FIRE survey from starburst to post-starburst. The survey obtained near-infrared spectra (0.82--2.51 \mic) of 93 local massive galaxies at different stages in the evolution from starburst to post-starburst, including systems with diverse emission line properties ranging from star-forming galaxies and AGN hosts, to traditional E+A galaxies with weak emission lines. These infrared spectra are supplemented by optical spectra from the \textit{Sloan Digital Sky Survey} (SDSS; \citealt{york00}), providing a spectral coverage from rest-frame 0.4 to 2.2 \mic, with two gaps in the infrared due to telluric absorption. 

In this paper, we collect photometry for the \textit{Sparks} galaxies, ranging from far-ultraviolet to far-infrared wavelengths. We fit the multi-wavelength SEDs and the optical spectra using the codes \texttt{Prospector} and \texttt{MAGPHYS} under a range of model assumptions. Our aim is twofold: (I) assess the robustness of derived parameters?such as stellar mass, SFR, and SFH parameters?for galaxies along the starburst-to-post-starburst sequence, particularly given the known systematic uncertainties between different SED fitting codes and the extreme properties that characterize this evolutionary phase, and (II) identify the parameters we will adopt as the final galaxy properties for the survey. 

The paper is organized as follows. In Section~\ref{sec:sample} we provide a brief description of the sample selection and main galaxy properties. In Section~\ref{sec:SEDs:photometry} we collect the multi-wavelength photometry, and in Section~\ref{sec:SEDs:modeling}, we describe our modeling approaches. Section~\ref{sec:results} presents our main results, where we compare between the physical properties derived using multi-wavelength SED fitting versus spectroscopic fitting (Section~\ref{sec:results:sed_fitting:spec-vs-sed}); compare between the properties derived using different codes and modeling assumptions (Section~\ref{sec:results:sed_fitting:different-codes}); and benchmark \texttt{Prospector}'s AGN torus model against AGN signatures in optical for the Sparks galaxies (Section~\ref{sec:results:sed_fitting:agn-torus}). We summarize in Section~\ref{sec:summary}.

Throughout the paper, we make use of $p$-values calculated for Pearson correlation coefficients and Kolmogorov--Smirnov (KS \citealt{press07}) tests. We consider $p$-value $< 0.01$ to be significant, and $p$-value $< 0.001$ to be highly significant.

\begin{figure*}
	\centering
\includegraphics[width=1\textwidth]{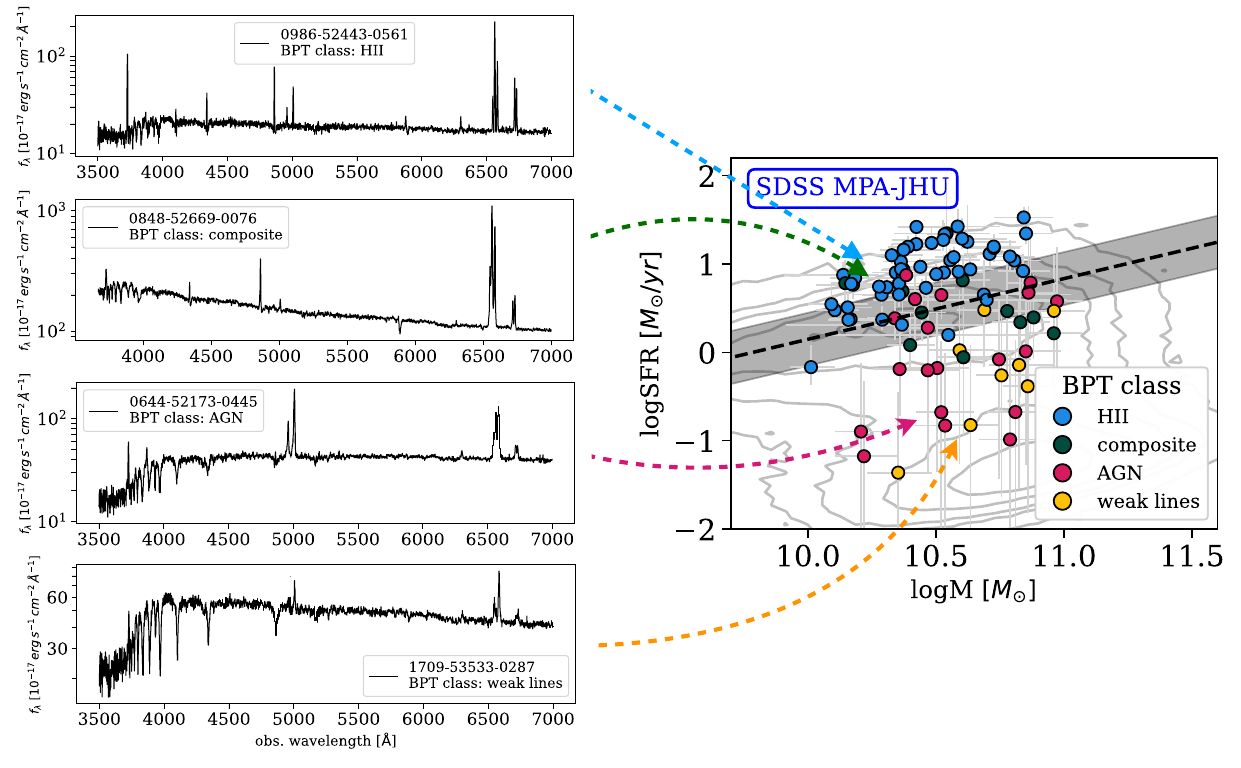}
\caption{\textbf{The \textit{Sparks} galaxy sample.} The right panel shows the Sparks galaxies in the SFR versus stellar mass plane, where both properties have been extracted from the MPA-JHU catalog. The galaxies have been selected in the stellar mass range $10^{10}$--$10^{11}\,M_{\odot}$, showing SFRs ranging across about three orders of magnitude with respect to the star-forming main sequence (black dashed line with $\pm 0.3$ dex as a gray band; based on \citealt{whitaker12}). The points are color-coded according to the galaxies' optical line ratio classifications. The gray background contours display all local SDSS galaxies for comparison. We emphasize that these properties have been used for sample selection, and the masses and SFRs have been revised according to the results of this work, shifting most of the composites and AGN hosts above the star-forming main sequence (see Paper I). The left panels show examples of four optical spectra from our sample, where the SDSS identifiers are the plate-MJD-fiberID of each source.}
\label{f:sample_selection_main_figure_paperII}
\end{figure*}

\section{Starburst to post-starburst sample}\label{sec:sample}

In this section, we give a brief overview of the sample selection and main properties of the \textit{Sparks} galaxies. A detailed description can be found in Paper I (Baron et al. submitted). Figure~\ref{f:sample_selection_main_figure_paperII} shows the galaxies in the SFR versus stellar mass plane, as well as four example optical spectra. Table~1 in Paper~I lists the galaxy coordinates and general properties.

The sample consists of 93 massive galaxies at $z \sim 0.1$ selected to span the starburst-to-post-starburst evolutionary sequence. The sample selection is based on the two-dimensional principal component analysis (PCA) space defined by \citet{wild07}, which uses the strength of the 4000 \AA\ break (PCA1) and the excess H$\delta$ absorption for a given 4000 \AA\ break strength (PCA2) to identify galaxies along the transition from starburst to post-starburst. Starting from all \textit{SDSS} DR7 galaxies with available PCA amplitudes in the redshift range 0.02--0.17, we selected extreme systems located in the tail of the two-dimensional distribution, applying different selection criteria across three bins to capture systems ranging from ongoing starbursts (bottom left of the PCA space) to aged post-starburst galaxies (upper region with PCA2 $>0$). Additional constraints included stellar masses in the range $10^{10}$--$10^{11}\,M_{\odot}$ from the MPA-JHU catalog \citep{kauff03b, b04, t04} and declinations below 20$^{\circ}$ for Magellan observability.

The sample has diverse optical emission line properties as classified by the MPA-JHU catalog: 53 galaxies (57\%) are classified as purely star-forming (`HII'), 11 (12\%) as composite systems showing both stellar and AGN ionization signatures, 21 (23\%) as AGN-dominated, and 8 (9\%) as weak-line systems. Initial stellar masses and star formation rates were adopted from the MPA-JHU catalog, which derived stellar masses from SPS fits to SDSS $ugriz$ photometry and estimated SFRs using either dust-corrected \halpha fluxes (for star-forming galaxies) or D$_n$4000 \AA\ indices (for composites, AGN, and weak-line systems where \halpha may not reliably trace star formation). The galaxies span approximately three orders of magnitude in star formation rate relative to the star-forming main sequence, from systems 1 dex above to 2 dex below the main sequence, providing comprehensive coverage of the transition from active star formation to post-starburst.

\section{Compilation of multi-wavelength SEDs}\label{sec:SEDs:photometry}

For each of the 93 \textit{Sparks} galaxies, we collect photometric measurements covering far ultraviolet to far-infrared wavelengths. Whenever possible, we use catalogs which performed forced photometry using the \textit{SDSS} source positions, as described below. We collect far and near ultraviolet photometry obtained with the \emph{The Galaxy Evolution Explorer} (GALEX; \citealt{martin05}). We use the improved \emph{GALEX} ultraviolet photometry by \citet{osborne23}, who performed forced model photometry using \emph{SDSS} priors on \emph{GALEX} imaging of $\sim$700\,000 galaxies included in the GALEX-SDSS-WISE Legacy Catalog (\emph{GSWLC}; \citealt{salim16}). The forced photometry on \emph{SDSS} source positions mitigates the impact of blending, which is of particular importance for this sample which includes merging galaxies with close companions. 

We use the $u$, $g$, $r$, $i$, and $z$-band \texttt{ModelMag} magnitudes from SDSS (\citealt{york00}). The \texttt{ModelMag} magnitudes yield the most accurate colors and model $\chi^{2}$ among the \emph{SDSS} photometric magnitudes (see section 3.1 in \citealt{salim16}). We use $J$, $H$, and $K$ near-infrared photometry from the \emph{2-Micron All-Sky Survey} (2MASS; \citealt{skrutskie06}). The \emph{SDSS} database provides astrometric cross-matches of objects observed by \emph{SDSS} and \emph{2MASS}, and we use the \emph{SDSS} {\sc casjobs}\footnote{\url{https://skyserver.sdss.org/CasJobs/default.aspx}} interface to extract the infrared photometry for each galaxy. Within \emph{2MASS}, some of the galaxies in our sample are resolved and have measurements in both the extended and point source catalogs, while others are not resolved and have measurements only in the point source catalog. Whenever a galaxy is spatially resolved in \emph{2MASS}, we use the magnitudes reported in the extended source catalog.

We use the W1--W4 near and mid-infrared photometry from the \emph{Wide-field Infrared Survey Explorer} (WISE; \citealt{wright10}). To mitigate blending in the WISE bands, we use the improved photometry by \citet{lang16}, who performed forced photometry using the \emph{SDSS} source positions in the \emph{unWISE} coaddds. The \citet{lang16} catalog lists the fraction of flux at the location of a source that is due to other sources. Within our sample, the 16th, 50th, and 84th percentiles of the fraction in W3 are $4.6 \times 10^{-5}$, $2.1 \times 10^{-4}$, and $7.7 \times 10^{-4}$, respectively, and the maximum fraction is 0.0039. The 16th, 50th, and 84th percentiles of the fraction in W4 are $1.8 \times 10^{-4}$, $5.5 \times 10^{-4}$, and $1.3 \times 10^{-3}$, respectively, with a maximum of 0.027. We manually inspected the imaging data of the 93 galaxies in our sample using the \emph{DESI} Legacy Image Viewer\footnote{\url{https://www.legacysurvey.org/viewer}} to confirm that there are no galaxies that are resolved in the \emph{SDSS} imaging but completely blended in \emph{WISE}. 
 
We collect far-infrared photometry at 60 and 100 $\mathrm{\mu m}$ from the \emph{Infrared Astronomical Satellite} (IRAS; \citealt{neugebauer84}). For each galaxy in the sample, we use {\sc scanpi} (\citealt{helou88}) to stack the calibrated survey scans around the \emph{SDSS} coordinates. We use the default settings of {\sc scanpi}, which are `Source Fitting Range'=3.2', `Local Background Fitting Range'=30', and `Source Exclusion Range for Local Background Fitting'=4' for the 60 \mic scans, and are `Source Fitting Range'=6.4', `Local Background Fitting Range'=30', and `Source Exclusion Range for Local Background Fitting'=6' for the 100 \mic scans (all values are given in arc minutes). In both cases, we use median stacks since the \emph{IRAS} data has non-Gaussian noise. The relevant {\sc scanpi} outputs include the best-fitting flux density ($f_{\nu}$), the RMS deviation of the residuals after the subtraction of the best-fitting template ($\sigma$), the offset of the peak of the best-fitting template from the galaxy coordinates ($\Delta$), and the correlation coefficient between the best-fitting template and the data ($\rho$). To avoid contamination by false matches and noise fluctuations, we consider a source detected if the following requirements are met: $\rho > 0.8$, $\Delta< 0.4'$, and $f_{\nu}/\sigma > 3$. These criteria were derived in \citet[see section 3.1 there]{baron22} and result in a contamination of $<1$\%. 

For sources that are considered detected by IRAS, we define the measured flux to be the best-fitting amplitude in the scan. For undetected sources, we define an upper limit on their flux value to be $3 \sigma$. The angular resolution of IRAS in 60 and 100 \mic is $1.5'$ and $2'$ (full width half maximum; FWHM). Our $\Delta< 0.4'$ criterion ensures that the peak emission is within 24\arcsec\, of the \emph{SDSS} source, minimizing the contamination from unrelated sources. However, since some of the galaxies in our sample are interacting, the large IRAS beam may include a contribution from the companion galaxy as well. We use the WISE W4 fluxes to derive correction factors for the far-infrared flux as described below. For each galaxy in the sample, we extract all the neighbors within 24\arcsec\, and use their W4 fluxes from \citet{lang16}. We define the far infrared correction factor: $\xi_{\mathrm{FIR}} = f_{s}(W4) / \big(f_{s}(W4) + \sum f_{n}(W4) \big)$, where $f_{s}(W4)$ is the W4 flux of the source, and $\sum f_{n}(W4)$ is the summed W4 flux of all the neighbors\footnote{We experimented with performing a weighted average that takes into account the distance of the W4 neighbors from the central source, and found comparable correction factors to those derived under a simple average. Since these corrections are much smaller than our adopted uncertainty, we choose to continue with the simple average. We get comparable correction factors when using W3 instead of W4.}. We then estimate the corrected far-infrared flux as $f_{corr}(\mathrm{FIR}) = \xi_{\mathrm{FIR}} \times f(\mathrm{FIR})$, where $\mathrm{FIR}$ stands for either 60 or 100 \mic. For upper limits, we do not correct the flux and keep the upper limit $f(\mathrm{FIR})$. If $f_{corr}(\mathrm{FIR}) < 3 \sigma$, we convert the measurement to be an upper limit, with a value of $f(\mathrm{FIR})$. If $f_{corr}(\mathrm{FIR}) \geq 3 \sigma$, we adopt $f_{corr}(\mathrm{FIR})$ as the $\mathrm{FIR}$ flux.

\begin{figure*}
	\centering
\includegraphics[width=1\textwidth]{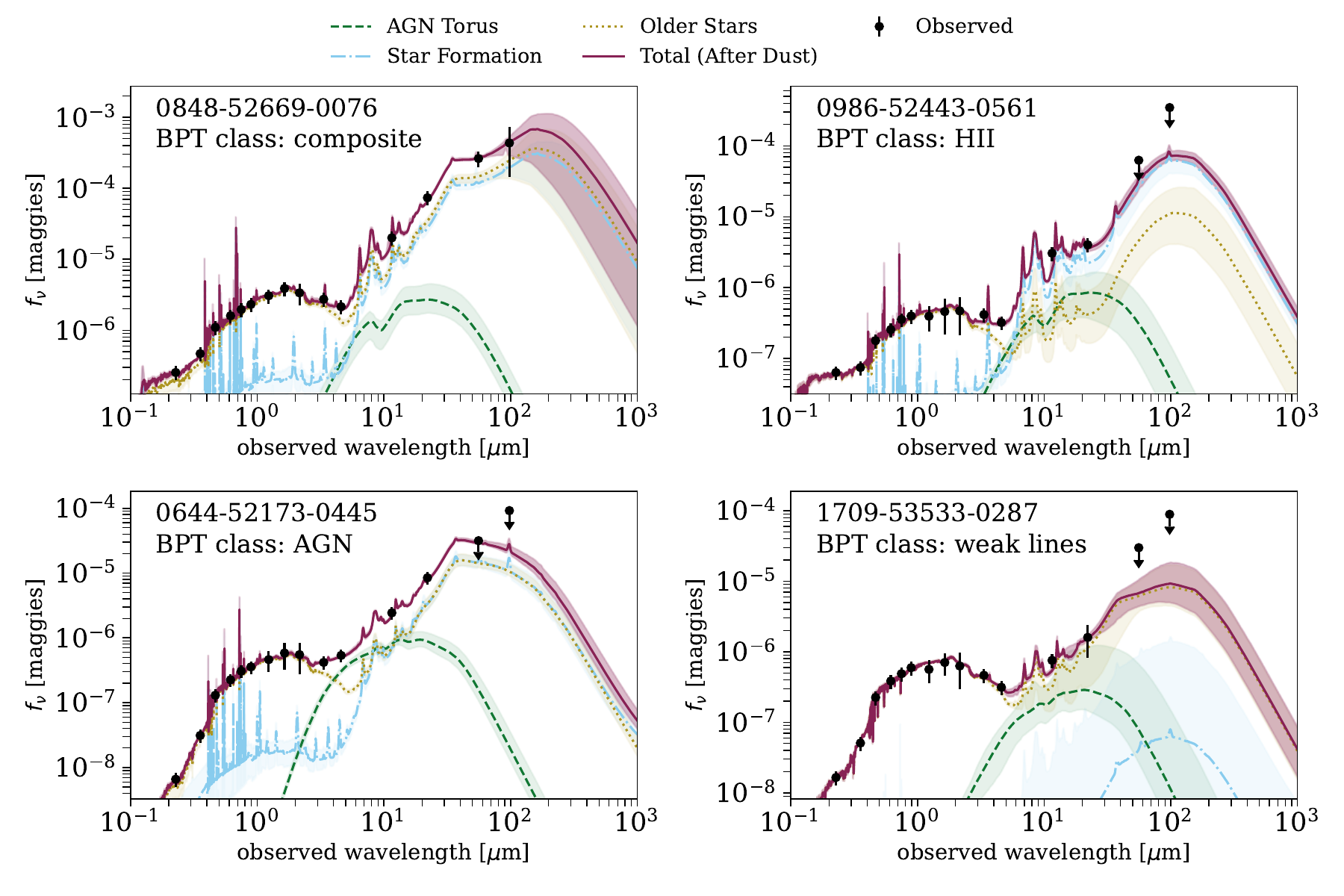}
\caption{\textbf{Example best-fitting SEDs obtained with \texttt{Prospector}.} The four panels show the best-fitting photometric-only SEDs produced using \texttt{Prospector} with the `free obscuration' model. Comparable fits are obtained with the `standard birth clouds' model. The sources are labeled with their SDSS identifiers (plate?MJD?fiberID) and SDSS optical line-ratio classifications. The black points show ultraviolet to far-infrared photometry (arrows represent upper limits), with vertical lines indicating $\pm 3 \sigma$ uncertainties. The different curves represent various components of the model, as indicated in the legend above. The fluxes are given in maggies, where 1 maggie corresponds to 3631 Jy.}
\label{f:SED_fitting_example}
\end{figure*}

Out of the 93 \textit{Sparks} galaxies, 47 are initially detected at 60\,\mic, 43 have upper limits, and 3 lack available scan data at their coordinates. At 100\,\mic, 17 galaxies are detected, 73 have upper limits, and 3 again lack scan coverage. Although some galaxies in the sample are interacting systems, manual inspection of their companions reveals that these are typically fainter by a factor of a few compared to the primary galaxy. Consequently, among the 93 galaxies, the far-infrared contamination parameter $\xi_{\mathrm{FIR}}$ is below 0.8 for 15 sources, and below 0.5 for a single source. As part of our adopted contamination correction, we reclassify 9 galaxies at 60\,\mic and 4 at 100\,\mic from detections to upper limits.

In all SED fits, we correct the IRAS photometry to an AB system, and account for the energy-counting nature of the detector via the filter definitions used to measure fluxes. We provide a table with the photometric measurements for all the sources online. 

\section{Stellar population synthesis modeling}\label{sec:SEDs:modeling}

We model the SEDs of our galaxy sample using two codes: \texttt{MAGPHYS} \citep{daCunha08, daCunha11} and \texttt{Prospector} \citep{leja17, johnson21}. These codes model the multi-wavelength emission from galaxies as a combination of ultraviolet-optical emission by stellar populations, dust attenuation and re-emission in infrared, and, when applicable, an AGN contribution. Both frameworks account for the energy balance between dust-attenuated stellar emission in the ultraviolet?optical and re-emission in the infrared, ensuring that the total absorbed energy equals the dust luminosity. Despite this shared principle, the codes differ in their treatment of stellar populations, star formation histories, AGN presence, and dust emission modeling, each having distinct regimes of applicability and limitations. 

\texttt{MAGPHYS} can be applied only to photometric observations, without explicit treatment of upper limits, while \texttt{Prospector} can fit spectroscopic and photometric observations simultaneously, taking into account upper limits. \texttt{MAGPHYS} models star formation histories (SFHs) using a parametric approach, combining an exponentially declining SFH with random bursts drawn from a fixed template library. In contrast, \texttt{Prospector} can fit the SFH using a non-parametric form that allows the star formation rate to vary freely across time bins. 

\begin{figure*}
	\centering
\includegraphics[width=1\textwidth]{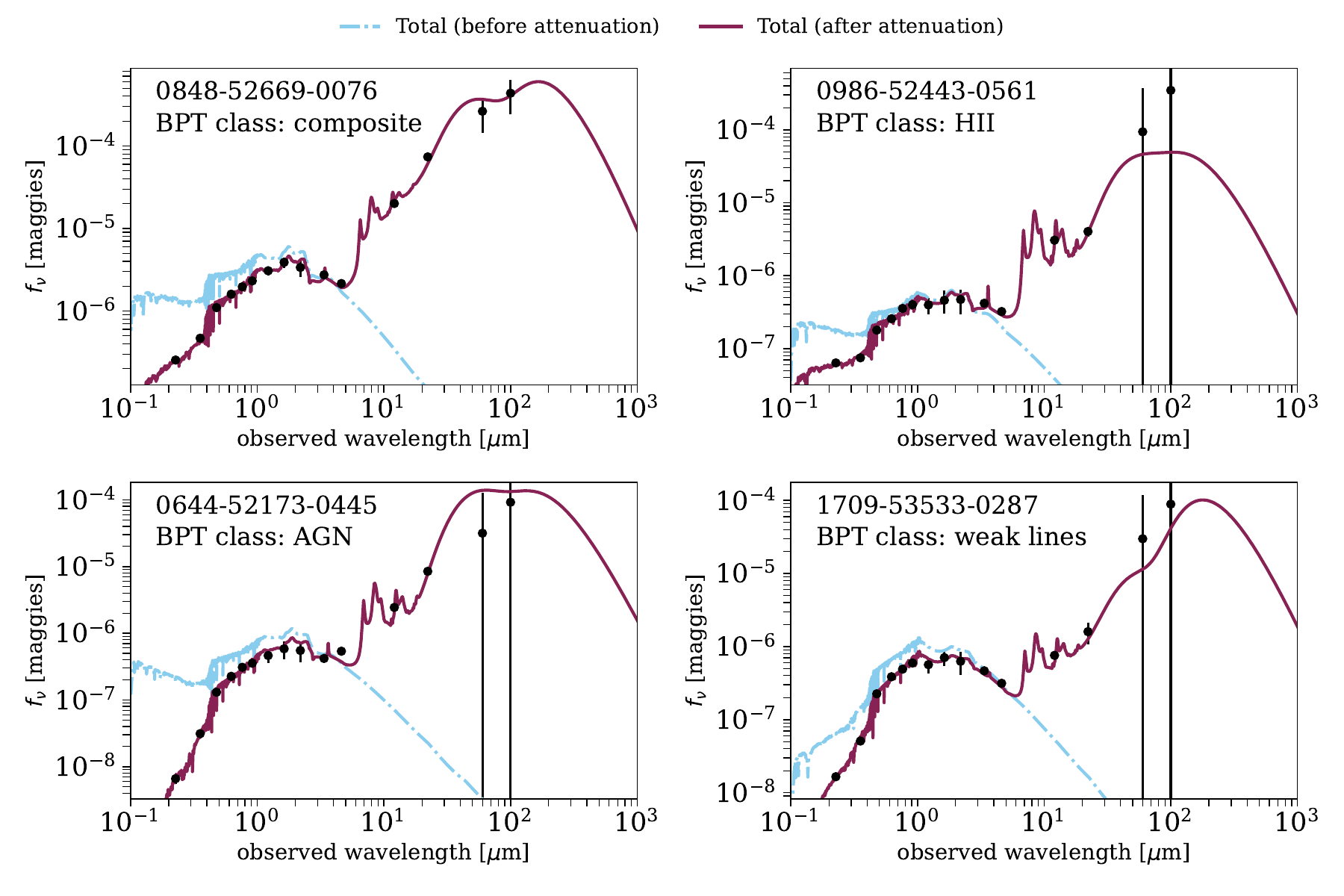}
\caption{\textbf{Example best-fitting SEDs obtained with \texttt{MAGPHYS}.} Best-fitting ultraviolet?to?far-infrared photometric SEDs produced with \texttt{MAGPHYS}. Each panel is labeled with the source?s SDSS identifier (plate?MJD?fiberID) and its SDSS optical line-ratio classification. Black points show the observed photometry, with vertical lines indicating $\pm 3\sigma$ uncertainties. Because \texttt{MAGPHYS} does not model upper limits, we treat these fluxes as detections with uncertainties set equal to the upper-limit value. The curves show the different model components, as indicated in the legend. The fluxes are given in maggies, where 1 maggie corresponds to 3631 Jy.}
\label{f:SED_fitting_example_magphys}
\end{figure*}

Both \texttt{MAGPHYS} and \texttt{Prospector} employ a two-component dust attenuation model that distinguishes between the dense birth clouds surrounding young stars and the diffuse ambient interstellar medium (ISM) affecting older stellar populations. In \texttt{MAGPHYS}, this approach follows the prescription of \citet{charlot00}: stars younger than 10 Myr experience attenuation from both the birth clouds and the diffuse ISM, whereas older stars are attenuated only by the ambient ISM. The attenuation curves and optical depths in \texttt{MAGPHYS} are drawn from a fixed library. In contrast, \texttt{Prospector} uses a similar age-dependent dust geometry but treats the optical depths and attenuation curves for both components as free parameters, often modeled with flexible parametric forms, which may represent different dust compositions or geometries with respect to the stars. 

For dust infrared emission, \texttt{MAGPHYS} uses a fixed set of empirically motivated templates using observations of the Milky Way or star-forming galaxies, with only a few parameters that are allowed to vary. PAH emission features, the near infrared continuum emission ($\mathrm{T_{d} = 850\, K}$), and the mid-infrared continuum emission ($\mathrm{T_{d} = 130\, K}$ and $\mathrm{T_{d} = 250\, K}$), are modeled using fixed templates with fixed spectral shapes, emissivity indices, and dust temperatures. It models grains in thermal equilibrium using three greybody components to represent warm grains ($\beta = 1.5$) in birth clouds and in the ambient ISM, and cold grains ($\beta = 2$) that reside only in the ambient ISM, allowing them to vary within a limited temperature range (30--60 K for warm grains, 15--25 K for cold grains). The fractional contributions of the different components to the infrared emission are fitted. In contrast, \texttt{Prospector} uses the physically motivated \citet{draine07} templates, with three free parameters: the PAH mass fraction ($q_{\mathrm{PAH}}$), the minimum radiation field intensity ($U_{\mathrm{min}}$), and the fraction of dust heated by a power-law distribution of stronger radiation fields ($\gamma_{e}$). The \citet{draine07} model assumes grain composition and size distribution that reproduce the average Milky Way extinction curve, and assumes that grains are heated by a scaled version of the \citet{mathis83} interstellar radiation field.

\texttt{MAGPHYS} does not include any contribution of AGN emission in its public version (see e.g., modification by \citealt{berta13}). While \texttt{Prospector} does model the mid-infrared emission by the AGN torus using a set of physically-motivated templates, the implementation we employ does not model the AGN contribution to ultraviolet-optical wavelengths. The inclusion of AGN torus mid-infrared emission therefore makes the implicit assumption that the AGN is fully obscured and that stellar continuum dominates in the rest-frame optical. For galaxies without an AGN, including a torus component violates \texttt{Prospector}?s energy-balance assumption, because the torus mid-infrared emission can vary freely without being tied to any ultraviolet?optical attenuation. Despite this limitation, we allow for an AGN component when fitting with \texttt{Prospector}, to test whether the emission in mid and far-infrared wavelengths in the AGN-dominated systems can be explained by an AGN.

For our sample, having four photometric measurements from WISE (3.4, 4.6, 12 and 22 \mic) and 0--2 photometric measurements from IRAS (60 and 100 \mic), we expect the dust infrared fitting to be under-constrained and to result in degeneracies between the different parameters. We use these fits only to study the total infrared luminosity, $L_{\mathrm{TIR}}$, and to test how well different SED fitting methods can reproduce the mid and far-infrared colors we observe.

In sections \ref{sec:SEDs:magphys} and \ref{sec:SEDs:prospector} below, we describe the specific options and settings used when running each code. Figure~\ref{f:SED_fitting_example} presents example best-fitting SEDs from \texttt{Prospector}, and Figure~\ref{f:SED_fitting_example_magphys} shows corresponding examples from \texttt{MAGPHYS}.

\subsection{\texttt{MAGPHYS} fits}\label{sec:SEDs:magphys}

We use the \texttt{MAGPHYS} package\footnote{\url{https://www.iap.fr/magphys/index.html}} described in \citet{daCunha08} with the latest filter file (last updated on March 2018). We use the \citet{bruzual03} SPS models.  We include all the photometric bands we collected: \emph{GALEX} far and near-ultraviolet; \emph{SDSS}  $u$, $g$, $r$, $i$, and $z$-band magnitudes; \emph{2MASS} $J$, $H$, and $K$-band magnitudes; \emph{WISE} W1, W2, W3, and W4; and \emph{IRAS} 60 \mic and 100 \mic. In the fitting, we assume the spectroscopic redshift from the \emph{SDSS}. 

\texttt{MAGPHYS} does not natively support the treatment of measurement upper limits. When excluding upper limits?particularly those from \emph{IRAS} at 60\,\micron\ and 100\,\micron?we found that the best-fitting SEDs often predicted fluxes exceeding these limits. To mitigate this, we included the upper limits in the \texttt{MAGPHYS} fits as if they were detections, assigning an uncertainty equal to the upper limit value. While this approach is not statistically equivalent to a proper upper limit treatment in the likelihood function (e.g., \citealt{isobe86}), we found that it effectively prevents the best-fitting SEDs from violating the upper limits in practice.

\subsection{\texttt{Prospector} fits}\label{sec:SEDs:prospector}

We perform two sets of fits using \texttt{Prospector}, one to the full panchromatic SED and the other to just the SDSS spectrum. In the spectroscopic fits, we fit only the SDSS spectrum, with no accompanying photometry. We also tried to perform a full spectro-photometric fit with all the available data, but were unable to obtain satisfactory fits with the current \texttt{Prospector} version (see Section~\ref{sec:results:sed_fitting:spec-vs-sed} for details). For the photometric fits, we include the same photometry as in the \texttt{MAGPHYS} fits. For both sets of fits, we use the dynesty dynamic nested sampling package \citep{speagle20}, the Flexible Stellar Population Synthesis (FSPS) models \citep{conroy09, conroy10}, the MILES spectral library \citep{vazdekis10}, and the MIST isochrones \citep{choi16, dotter16}. We assume a \cite{chabrier03} initial mass function and fix the model redshift to the spectroscopic redshift. Before fitting the spectroscopic data, we convolve all models to the \textit{SDSS} wavelength-dependent spectral resolution. 

\begin{figure*}
	\centering
\includegraphics[width=1\textwidth]{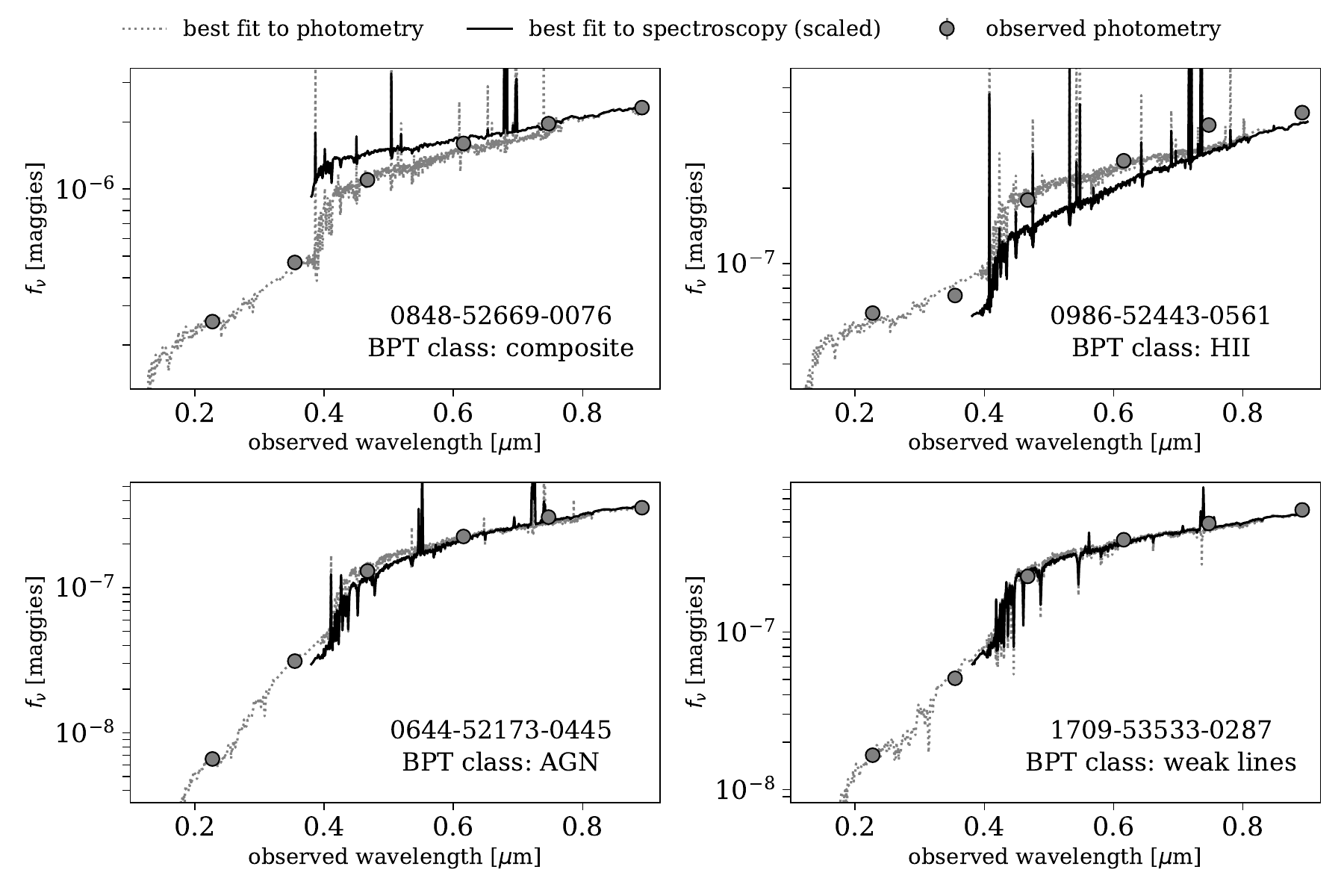}
\caption{\textbf{Comparison of \texttt{Prospector} fits to SDSS spectroscopy and multi-wavelength photometry.} Each panel shows the best-fitting spectrum over the ultraviolet?optical range, modeled under the standard birth-cloud assumption. Photometric measurements are plotted as gray points, with their best-fitting model shown as a dotted gray line. The best fit to the SDSS spectrum is shown in black and rescaled to match the photometric flux at 9000 \AA, allowing a direct comparison of spectral shapes within the fiber and for the full galaxy. For the two galaxies in the top row, clear color differences are visible between the fiber spectrum and the integrated photometry.}
\label{f:SED_and_spec_fitting_example}
\end{figure*} 

For both sets of fits, we assume a non-parametric star formation history using the continuity prior \cite{leja19}, with the logarithmic ratio between neighboring bins sampled from a Student's t distribution centered at 0, with \texttt{scale=0}, and $\nu=1$. We set the three most recent bins to span 10, 40, and 50 Myr, and the remaining bins logarithmically scan the rest of the age of the universe. For the spectroscopic fit, we utilize 10 total bins, and for the photometric fit we use 7. 

We allow the stellar mass to vary over $\log(M_{\star,\mathrm{formed}}/M_\odot)\in[7,12]$ with a uniform prior in log space, and the stellar velocity dispersion over $\sigma_*/\mathrm{km,s^{-1}}\in[0,400]$ with a uniform prior in linear space. The metallicity is sampled in log space, with a clipped normal distribution prior centered at 0 with $\sigma=0.25$, with $\log(Z)\in[-1.0, 0.19]$. We assume the \cite{kriek16} dust law with a free dust index $n_{dust}\in[-1, 0.4]$. The mid to far-infrared dust emission shape is governed by the \cite{draine07} dust models, sampled logarithmically with $U_{min}\in[0.1, 25]$, $q_{\mathrm{PAH}}\in[0.1, 10]$, and $\gamma_e\in[0.001, 1]$, while enforcing energy balance.

We adopt the \cite{charlot00} prescription where old ($t>10$ Myr) stars are exposed to a dust screen of optical depth $\tau_{old}$ and young stars ($t<10$ Myr) are exposed to an additional screen that represents their birth clouds, with $\tau_{young}=\tau_{BC} + \tau_{old}$. We allow $\tau_{old}$ to vary in the range [0, 2.3] (corresponding to a maximum $A_v=2.5$). We run two flavors of models with regard to $\tau_{BC}$: (i) `standard birth clouds': we use the standard assumption $\tau_{BC}=\tau_{old}$, where the attenuation of the young stellar population is double that of the old stars (e.g., \citealt{wild20}), and (ii) `free obscuration': we leave $\tau_{BC}$ free, with a prior centered at 1.5 with $\sigma=1.8$, truncated between 0 and 10 (designed to approximate the \texttt{MAGPHYS} $\tau_{BC}$ prior in a functional form that can be sampled in dynesty nested sampling). For the spectroscopic fits where we do not have far-infrared data to constrain the total infrared luminosity, we only fit with the $\tau_{BC}=\tau_{old}$ prior.

In addition to the far-infrared emission linked to the attenuated ultraviolet stellar radiation, the model for the photometric fits also includes AGN mid-infrared torus emission using the \cite{nenkova08} models as implemented in \cite{leja19}. These models parameterize the AGN emission via two free parameters, $f_{AGN}$, the fraction of the AGN luminosity to the total bolometric luminosity of the starlight, and $\tau_{AGN}$, the optical depth of the torus. We follow \cite{leja19b} and allow these to vary in the range $f_{AGN}\in[10^{-5}, 3]$ and $\tau_{AGN}\in[5, 150]$. Since \texttt{Prospector} makes the assumption that the AGN is fully obscured in optical wavelengths, such a component is not included in the fits to the optical spectroscopy of the stellar continuum. 

In the photometric fit, nebular emission is left on, with \texttt{gas\_logu} free $\in[-4,-1]$. In the spectroscopic fit, we fit the optical emission lines using Gaussian profiles centered on the lines with a free width in the range 0--400 $\mathrm{km\,s^{-1}}$. Since the sources in sample have diverse ionization properties (dominated by star formation, AGN, or shocks), we do not tie the line luminosities to the best-fitting stellar continua. We handle this fitting through the \texttt{Prospector} nebular marginalization routine, which finds the best fitting Gaussian for each line after subtracting the stellar continuum during each iteration before measuring the likelihood. Finally, we allow the standard \texttt{Prospector} spectroscopic jitter error inflation term (in the range [0.5,3.0]) and spectroscopic outlier fraction term (in the range [$10^{-5}$,0.2]).

\section{Results}\label{sec:results}

In section \ref{sec:results:sed_fitting:spec-vs-sed} we compare between the physical parameters derived from fitting the rest-frame optical spectra versus the far-ultraviolet to far-infrared photometry, within a single controlled framework, with \texttt{Prospector}. In section \ref{sec:results:sed_fitting:different-codes} we benchmark \texttt{MAGPHYS} and \texttt{Prospector} and compare their outputs with SDSS-derived properties (MPA-JHU catalog; \citealt{b04, kauff03b, t04}) which include additional information not included in the \texttt{MAGPHYS} and \texttt{Prospector} fits (H$\alpha$ flux and optical line ratios). For \texttt{Prospector}, we include fits based on the spectroscopy, and two fits based on the multi-wavelength SEDs where we vary our assumptions on dust attenuation. In Section~\ref{sec:results:sed_fitting:agn-torus}, we characterize \texttt{Prospector}'s torus model, by benchmarking it against AGN signatures in optical, and comparing the derived torus properties to those of mid-infrared-selected AGN.

\begin{figure*}
	\centering
\includegraphics[width=1\textwidth]{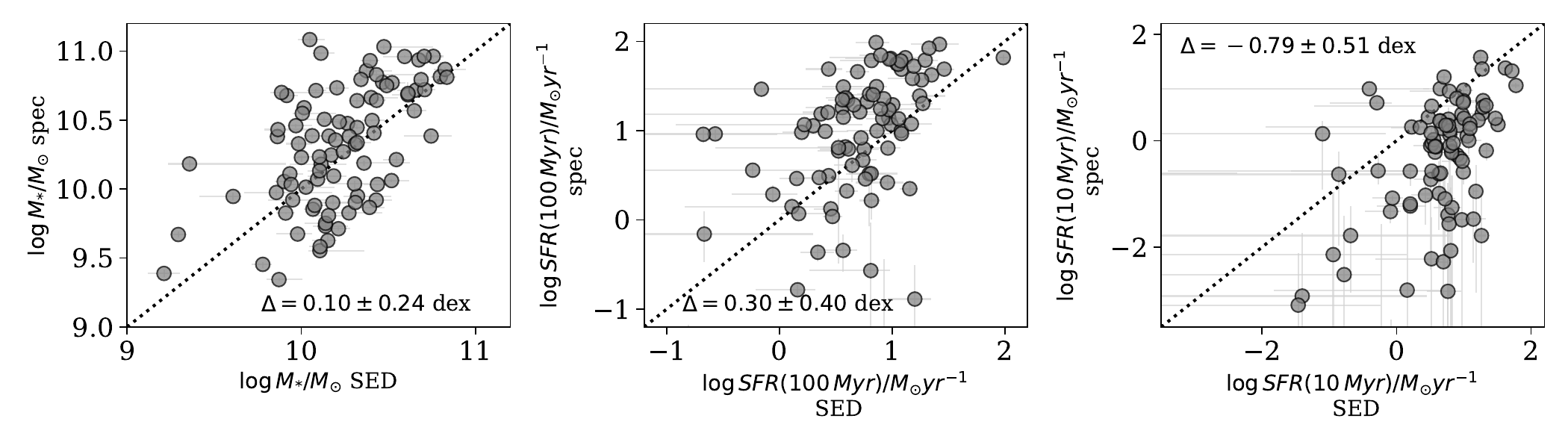}
\caption{\textbf{Comparison of best-fitting stellar masses and SFRs for the spectroscopic and photometric \texttt{Prospector} fits.} The panels show the stellar mass, the average SFR over the last 100 Myrs, and over the last 10 Myrs. The black dashed lines mark the 1:1 relation. In each panel, $\Delta$ denotes the median and median absolute deviation (MAD) of the logarithmic difference between the properties derived into the two fits.}
\label{f:Pros_SED_vs_spec_comparison}
\end{figure*}

\subsection{Comparison of spectroscopic and photometric SED fits by \texttt{Prospector}}\label{sec:results:sed_fitting:spec-vs-sed}

Rest-frame optical spectra and multi-wavelength photometry offer distinct views on the stellar populations, gas, and dust in galaxies that transition from starburst to post-starburst. Far-ultraviolet to far-infrared photometry captures a larger fraction of the radiation originating from stellar populations, as well as radiation absorbed and re-emitted in infrared by dust. As a result, it provides a better leverage on the total stellar mass, dust attenuation, and dust-obscured star formation. However, as it averages over different stellar populations, it might be less sensitive to short-lived post-burst features. In contrast, the optical spectrum traces the D$_{n}$4000\AA~ break and strong Balmer absorption features, which are produced by A-type stars with lifetimes of a few hundred million to a billion years. It is therefore expected to provide a more precise constraint on the SFH over the past $\sim$1 Gyr. With no direct constraint on the near and far infrared emission, SPS fitting of the optical continuum may suffer from large systematic uncertainties on the derived stellar mass, and can  underestimate the amount of recent heavily obscured star formation.

The most precise constraints on the physical properties of starburst and post-starburst galaxies will most likely come from a fit that combines the optical spectrum with the multi-wavelength photometry. While \texttt{Prospector} can jointly model spectroscopic and photometric observations, we found the framework at its current stage to be insufficient for modeling the \textit{Sparks} galaxies. Since the optical spectra were obtained using 3\arcsec~ fibers which do not cover the galaxies entirely, there are differences in colors between the fiber and total galaxy band magnitudes (see figure \ref{f:SED_and_spec_fitting_example}). These differences can be as large as $\pm 0.2$ dex in $(u-g)$ and $(g-r)$ colors (see appendix \ref{app:sed_extra:pros-spec-photo}). While \texttt{Prospector} does have the ability to account for aperture effects via the application of a polynomial calibration vector, the application of such a method to these data is uncomfortable because the deviation between the spectroscopic and photometric shape is not driven by slit losses, but instead by physical color gradients within galaxies. 


An additional challenge is the large statistical imbalance between the spectroscopic and photometric information. With $\sim$1000 statistically independent spectroscopic measurements compared to 15 photometric measurements, the fit reproduced the spectrum well while being heavily-influenced by the prior in infrared wavelengths, underestimating the emission at 60 and 100 \mic by 0.3--1 dex. Although this statistical imbalance could be partially mitigated by artificially increasing the uncertainties of the spectroscopic data points, this ad-hoc re-weighting proved insufficient to produce a satisfactory joint fit in many of the systems. 

A joint photometric and spectroscopic fit for the \textit{Sparks} galaxies requires new routines within \texttt{Prospector} to account for aperture mismatches and to  properly balance disparate data types, which are beyond the scope of this paper. Due to the limitations in joint-fitting, we proceed by analyzing the spectroscopic and photometric datasets independently, treating them as two distinct but complementary probes of the galaxies' physical properties. As discussed in Section~\ref{sec:SEDs:prospector}, the \texttt{Prospector} spectroscopic fits treat the luminosities of nebular lines such as \halpha as independent parameters rather than tying them to the stellar continuum. This approach is necessary because 40 of the 93 \textit{Sparks} galaxies exhibit optical line ratios inconsistent with ionization by young, massive stars (composites, AGN, and LINERs; see Paper~I). As a result, the inferred SFHs are constrained solely by the stellar continuum emission, and not by the observed \halpha flux. The best-fitting properties presented in this section are based on fits under the standard birth-clouds assumption using similar priors for the dust attenuation and SFH, as described in section \ref{sec:SEDs:prospector}. Figure \ref{f:SED_and_spec_fitting_example} shows example ultraviolet?optical spectra from the two best-fitting models, one based on spectroscopy and the other on photometry. The best-fitting spectra are normalized to have same flux at 9000 \AA, illustrating the difference in colors traced within and outside the SDSS fiber.

In figure \ref{f:Pros_SED_vs_spec_comparison} we compare stellar masses and average SFRs over the past 100 Myr and 10 Myr, derived separately from spectroscopic and photometric fits. The spectroscopic values are scaled by the SDSS aperture correction factors for mass and SFR from the MPA-JHU catalog\footnote{We tested the relations without applying SDSS aperture corrections. In all three properties, omitting the correction changes the overall bias--as the stellar mass and SFR within the fiber are smaller--but yields comparable scatters. We therefore conclude that the scatter is not driven by the aperture corrections.}. Because the photometry includes near-infrared coverage, we expect it to provide more reliable estimates of the stellar mass. For systems hosting obscured star formation, we expect the far-infrared photometry to provide further constraints on the SFR. 

We further note that the fits to the photometry include an AGN mid-infrared torus component, which could, in principle, account for mid-infrared emission that would otherwise be attributed to star formation, leading to lower derived SFRs compared to the spectroscopic fits. We do not expect this to have a significant impact since: (i) the AGN torus emission peaks in mid-infrared wavelengths while the radiation by dust heated by star formation peaks at far-infrared wavelengths (e.g., figure \ref{f:SED_fitting_example}), and (ii) even when including an AGN torus component, its fractional contribution to the mid-infrared emission is $<0.2$ for most of the galaxies (see section \ref{sec:results:sed_fitting:agn-torus}). In practice, as we discuss below, we find the opposite: higher derived instantaneous SFRs for the photometric fits compared to the spectroscopic ones. This is discussed further in section \ref{sec:results:sed_fitting:different-codes}, where we compare \texttt{Prospector} to \texttt{MAGPHYS} fits to the photometry, where the latter does not include an AGN torus component.

The larger error bars in figure \ref{f:Pros_SED_vs_spec_comparison} for photometry-based estimates reflect the smaller number of statistically independent measurements used in the fit (15 versus $\sim$1000 for spectra). However, since error bars do not account for model mis-specification, particularly relevant when deriving integrated properties from optical-only spectra, we suggest that the uncertainties on spectroscopy-based properties are underestimated.

\begin{figure}
	\centering
\includegraphics[width=\columnwidth]{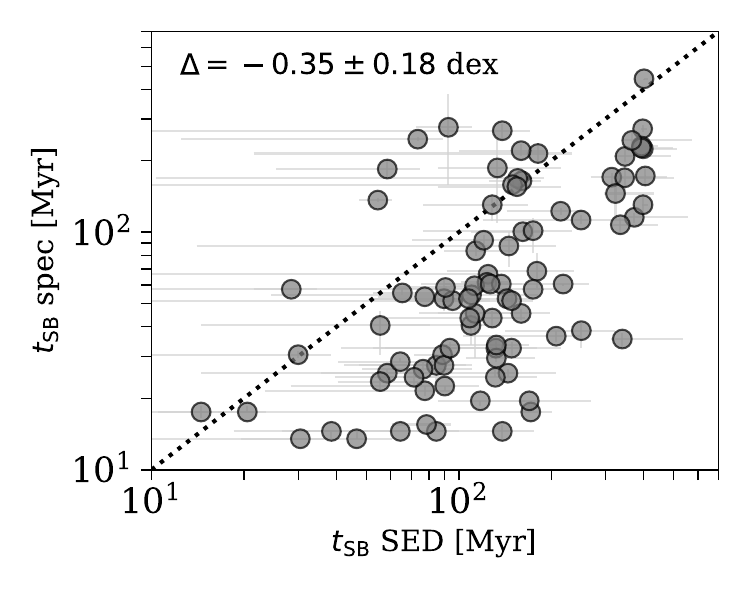}
\caption{\textbf{Comparison of starburst ages from spectroscopic and photometric \texttt{Prospector} fits.} Gray points show ages derived from the best-fitting non-parametric SFHs using optical spectroscopy versus multi-wavelength photometry. Spectroscopy provides tighter constraints on the SFH over the past $\sim$1 Gyr due to strong Balmer absorption features. The black dashed line marks the 1:1 relation, and $\Delta$ indicates the median and MAD of the logarithmic difference. }
\label{f:PSB_Prospector_comparison}
\end{figure}

Despite adopting similar priors and model assumptions, we find differences in all three derived properties. The median stellar-mass offset between spectroscopic and photometric fits is 0.1 dex, with a scatter of 0.24 dex (median absolute deviation; MAD) around the 1:1 relation. Relative to photometry, the spectroscopic fits yield SFRs that are higher by 0.3 dex over the past 100 Myr and lower by 0.8 dex over the past 10 Myr. The direction of this bias is expected: strong Balmer absorption lines in the optical spectra drive the fits toward a rapidly declining SFH, an effect less pronounced in broadband photometry.

For comparison, we find a much better agreement when comparing the properties derived by fitting the photometry alone, for the standard birth-clouds and free obscuration models. We find median differences of 0.03--0.05 dex, and scatters around the 1:1 relations of 0.06-0.14 dex (MAD; see figure \ref{f:Pros_SED_comparison_fixed_vs_free} in the appendix). 

In figure \ref{f:PSB_Prospector_comparison} we compare the starburst age $t_{\mathrm{SB}}$ derived from the best-fitting SFH for the spectroscopic and photometric fits. In this work, $t_{\mathrm{SB}}$ is defined as the lookback time when 90\% of the stellar mass formed within the last 1~Gyr. Because spectroscopy captures strong Balmer absorption features from intermediate-age stars, we expect it to provide a more precise constraint on the SFH over the past $\sim$1 Gyr. As expected, the photometric fits yield larger uncertainties due to the smaller number of fitted data points and the lower sensitivity of broad-band SEDs to recent bursts. Even with the larger uncertainties, we find systematic differences between the two methods: spectroscopy generally yields younger $t_{\mathrm{SB}}$ values, with 62 galaxies having $t_{\mathrm{SB}} < 100$ Myr compared to only 33 in the photometric fits.

What drives the large differences in derived properties between the spectroscopic and photometric fits? A natural hypothesis is aperture mismatch: the fiber samples only the central regions of each galaxy, which may host different stellar populations than the outskirts. To test this, we color-coded the points presented in figures \ref{f:Pros_SED_vs_spec_comparison} and \ref{f:PSB_Prospector_comparison} by the differences between fiber and total broad-band colors, $(u-g)$, $(g-r)$, and $(r-i)$, and show these in \ref{app:sed_extra:pros-spec-photo}. The $(u-g)$ color is most sensitive to stellar population differences, while $(r-i)$ helps disentangle dust reddening from population effects.

\begin{figure}
	\centering
\includegraphics[width=\columnwidth]{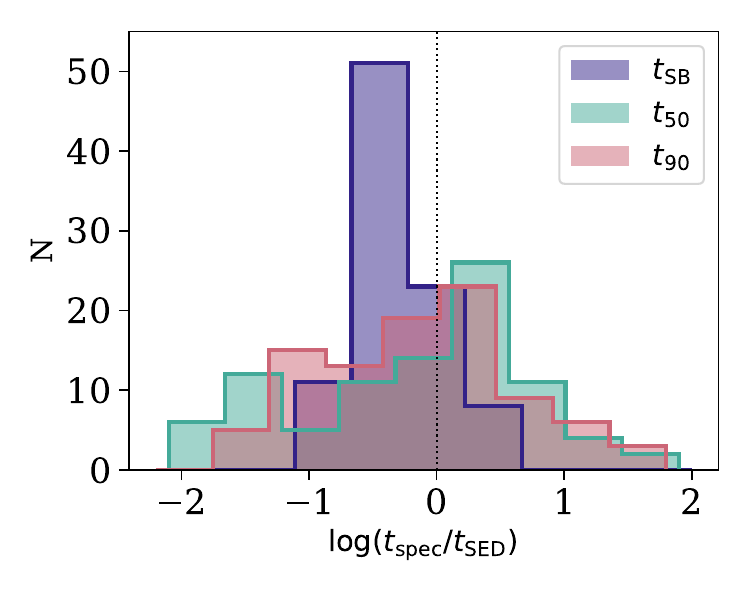}
\caption{\textbf{Comparison of best-fitting SFHs from spectroscopic and photometric \texttt{Prospector} fits.} The histograms show the logarithmic differences in timescales derived from the SFHs in the two cases. The starburst age is denoted $t_{\mathrm{SB}}$, while $t_{50}$ and $t_{90}$ indicate the lookback times when 50\% and 90\% of the stellar mass was formed. Differences of up to 2 dex in $t_{50}$ and $t_{90}$, and up to 1 dex in $t_{\mathrm{SB}}$, reveal substantial discrepancies between spectroscopy- and photometry-derived SFHs.}
\label{f:Pros_SED_vs_spec_comparison_timescales}
\end{figure}

Figures \ref{f:Pros_SED_vs_spec_comparison_band_colors} and \ref{f:PSB_Prospector_comparison_colors} in \ref{app:sed_extra:pros-spec-photo} show that color differences inside versus outside the fiber do not drive the observed discrepancies in the derived properties. For each property, we define $\Delta y = \log(x_{\mathrm{spec}}) - \log(x_{\mathrm{SED}})$, where $x$ is stellar mass, SFR, or $t_{\mathrm{SB}}$, and compute the Pearson correlation with $(u-g)_{\mathrm{fiber}} - (u-g)_{\mathrm{total}}$ and the other color gradients. No significant correlations are found. We therefore conclude that aperture mismatch does not drive the observed differences in derived properties. 

\begin{figure*}
	\centering
\includegraphics[width=1\textwidth]{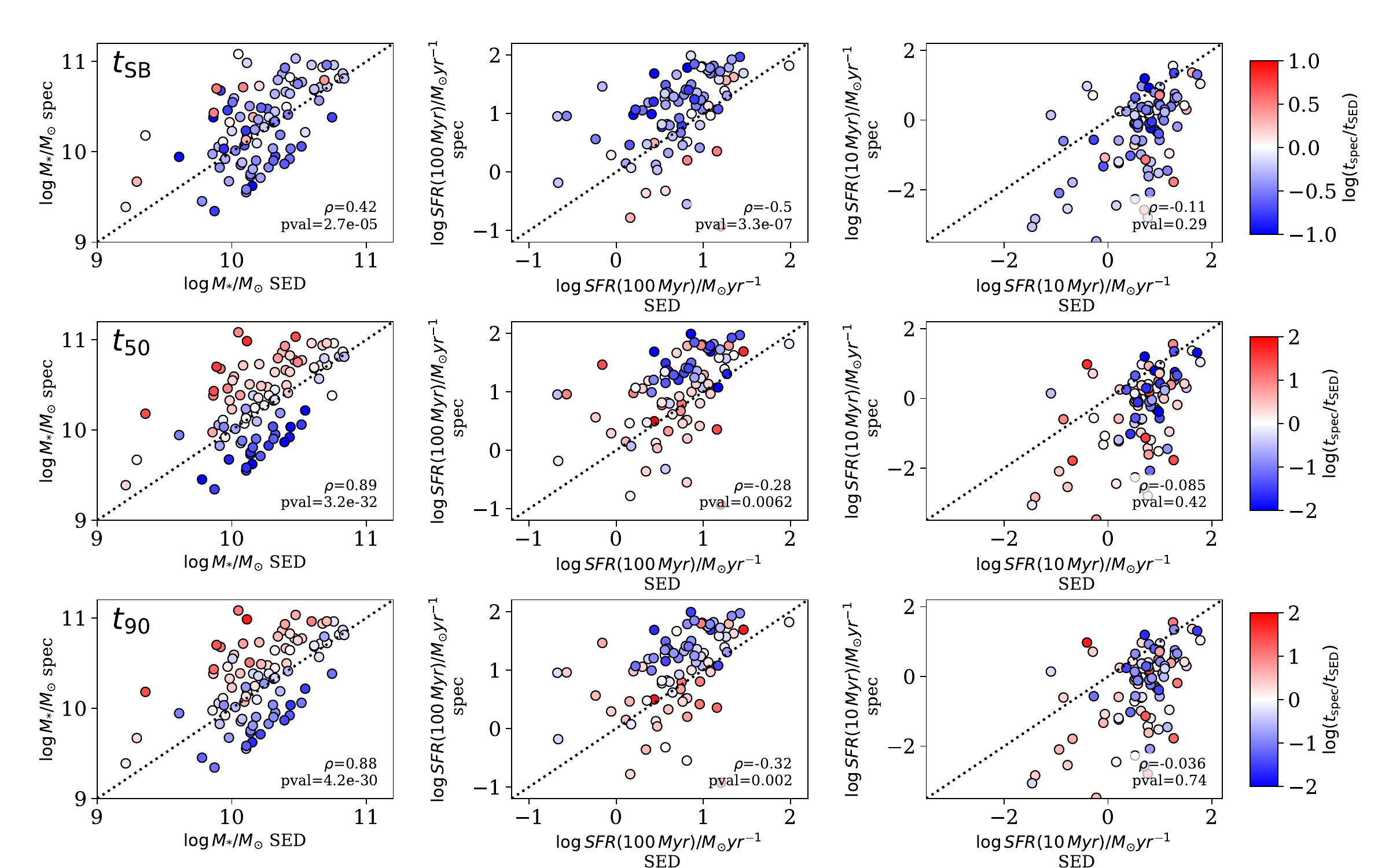}
\caption{\textbf{Differences in derived physical properties between spectroscopy and photometry are linked to differences in the SFH.} As in figure \ref{f:Pros_SED_vs_spec_comparison}, the panels compare stellar mass and average SFRs over the last 100 Myr and 10 Myr from spectroscopic versus photometric fits. Points are color-coded by the differences in derived timescales $t_{\mathrm{SB}}$, $t_{50}$, and $t_{90}$. In each panel, $\rho$ and $pval$ denote the Pearson correlation coefficient and its p-value between $\Delta y = \log(x_{\mathrm{spec}}) - \log(x_{\mathrm{SED}})$ (where $x$ is stellar mass or SFR) and $\log(t_{\mathrm{spec}}/t_{\mathrm{SED}})$. Clear color gradients in several panels indicate strong and significant correlations between the physical properties and the best-fitting SFH.}
\label{f:Pros_SED_vs_spec_comparison_SFH}
\end{figure*}

The different starburst ages from the spectroscopic and photometric fits reflect different best-fitting SFHs over the past 1 Gyr. To compare between the full SFHs, figure \ref{f:Pros_SED_vs_spec_comparison_timescales} shows the distributions of differences in timescales derived from the SFHs. In addition to the starburst age, we compare $t_{50}$ and $t_{90}$, the lookback times when 50\% and 90\% of the stellar mass was formed. Differences of up to 2 dex in $t_{50}$ and $t_{90}$, and up to 1 dex in $t_{\mathrm{SB}}$, highlight how spectroscopy and photometry can paint very different pictures of these galaxies? evolution.

In figure \ref{f:Pros_SED_vs_spec_comparison_SFH} we show that the discrepancies in stellar mass and SFR(100 Myr) arise directly from the different best-fitting SFHs. The stellar mass differences are strongly correlated with differences in $t_{50}$ and $t_{90}$, and more weakly with $t_{\mathrm{SB}}$. The sign of these correlations follows the expected variation in mass-to-light ratio with age: when spectroscopy implies earlier star formation, $t_{90}(\mathrm{spec}) > t_{90}(\mathrm{SED})$, the fitted populations are older and dimmer, requiring a higher stellar mass to reproduce the observed flux. Because most of the mass formed before the recent or ongoing burst, the correlations with $t_{50}$ and $t_{90}$ dominate. The differences in SFR(100 Myr) show the opposite trend, correlating negatively with $t_{\mathrm{SB}}$--again consistent with expectations from age-dependent mass-to-light ratios. We do not find strong and significant correlations between differences in SFR(10 Myr) and any of the timescales.

\begin{figure*}[t!]
	\centering
\includegraphics[width=0.9\textwidth]{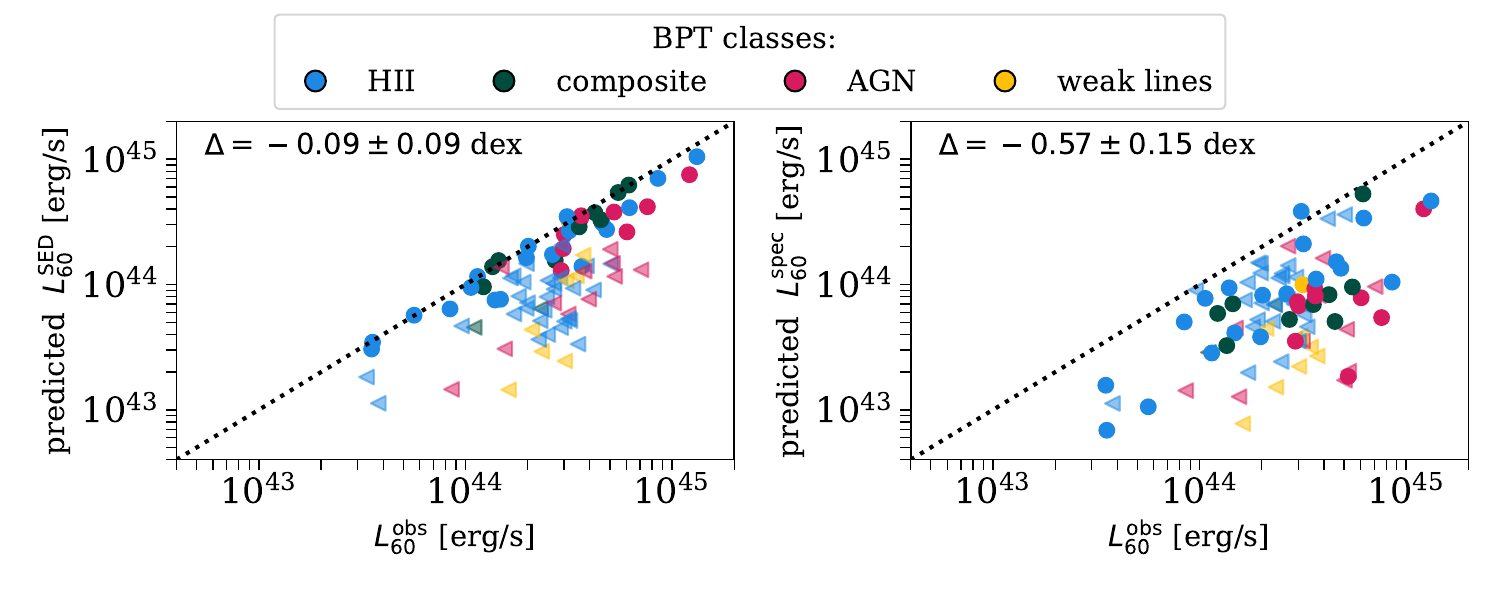}
\caption{\textbf{Predicted versus observed $L$(60 \mic) from photometric and spectroscopic \texttt{Prospector} fits.} The panels show the predicted 60 \mic\ luminosity from the multi-wavelength SED fit (left) and the optical spectroscopy fit (right). IRAS detections at 60 \mic\ are marked with circles, and upper limits with left-pointing triangles. The markers are color-coded according to the MPA-JHU classification of the source, as indicated in the legend above. The black dashed line marks the 1:1 relation, and $\Delta$ gives the median and MAD of the logarithmic difference. Without far-infrared information, the spectroscopic fit systematically underestimates the 60 \mic\ emission.}
\label{f:Pros_comparison_spec_vs_SED_FIR}
\end{figure*}

Figure \ref{f:Pros_comparison_spec_vs_SED_FIR} compares the predicted 60 \mic\ luminosity from the multi-wavelength SED and optical spectroscopy fits to the observed values. Without infrared data, the spectroscopic fit underestimates $L_{60}$ by a large factor (0.57 dex), which likely contributes to the large differences in derived SFR(10 Myr) between spectroscopic and photometric fits. The points in the figure are color-coded according to the MPA-JHU classification of the source, with the classes displaying the following median and MAD offsets: (-0.48, 0.22) for the star-forming galaxies, (-0.71, 0.09) for the composites, (-0.77, 0.16) for the AGN hosts, and (-0.5, 0.1) for the weak-line galaxies. We attribute the offset observed for star-forming galaxies to the spectroscopic fit being driven by the strong Balmer absorption features which prefers intermediate-age stellar populations at the expense of ongoing star formation. This is also supported by our findings in Section~\ref{sec:results:sed_fitting:different-codes} that the spectroscopic fits underpredict the SFR as derived from \halpha flux. The additional offset seen in AGN hosts relative to star-forming galaxies may arise from the absence of an AGN mid-infrared torus component in the spectroscopic \texttt{Prospector} model.

Figure~\ref{f:Pros_comparison_spec_vs_SED_FIR} shows that the photometric fits also slightly underestimate $L_{60}$, by 0.09 dex. As we describe in detail in section \ref{sec:results:sed_fitting:different-codes} below, using \texttt{Prospector}, we were unable to fit the SED without underestimating the far infrared by some factor, even when changing the dust attenuation model or model priors. We encountered the same issue when jointly fitting the spectroscopy and photometry as well, with differences of 0.3--1 dex between observed and predicted $L$(60 \mic).

\textbf{In summary}, \texttt{Prospector} spectroscopic fits to the stellar optical continuum and photometric fits to the panchromatic SED produce systematically different galaxy properties. Spectroscopic fits typically infer younger $t_{\mathrm{SB}}$ and higher $\mathrm{SFR}(100\,\mathrm{Myr})$, but lower $\mathrm{SFR}(10\,\mathrm{Myr})$, relative to photometric fits. These differences arise because the spectroscopic fits are more strongly constrained by Balmer absorption lines, which favor rapidly declining SFHs resulting in intermediate-age stellar populations. Differences in stellar mass and SFR(100 Myr) correlate strongly with discrepancies in SFH timescales ($t_{50}$ and $t_{90}$) rather than fiber-to-total color differences, indicating that the best-fitting SFH is the main source of scatter. The best-fitting models to the optical continuum emission underpredict the expected far-infrared luminosity, and in Section~\ref{sec:results:sed_fitting:different-codes} below, we also show that they underpredict the SFR as derived from \halpha flux. Therefore, in our sample, SPS fitting of the optical continuum emission misses ongoing star formation, which may lead to biases in the derived starburst ages. Photometric fits better capture the total stellar mass and star formation, but their lower sensitivity to short-lived bursts results in larger uncertainties in $t_{\mathrm{SB}}$. Overall, robust estimates of $t_{\mathrm{SB}}$ likely require either far-infrared information or improved joint spectro-photometric modeling that accounts for aperture effects.

\begin{figure*}
	\centering
\includegraphics[width=0.7\textwidth]{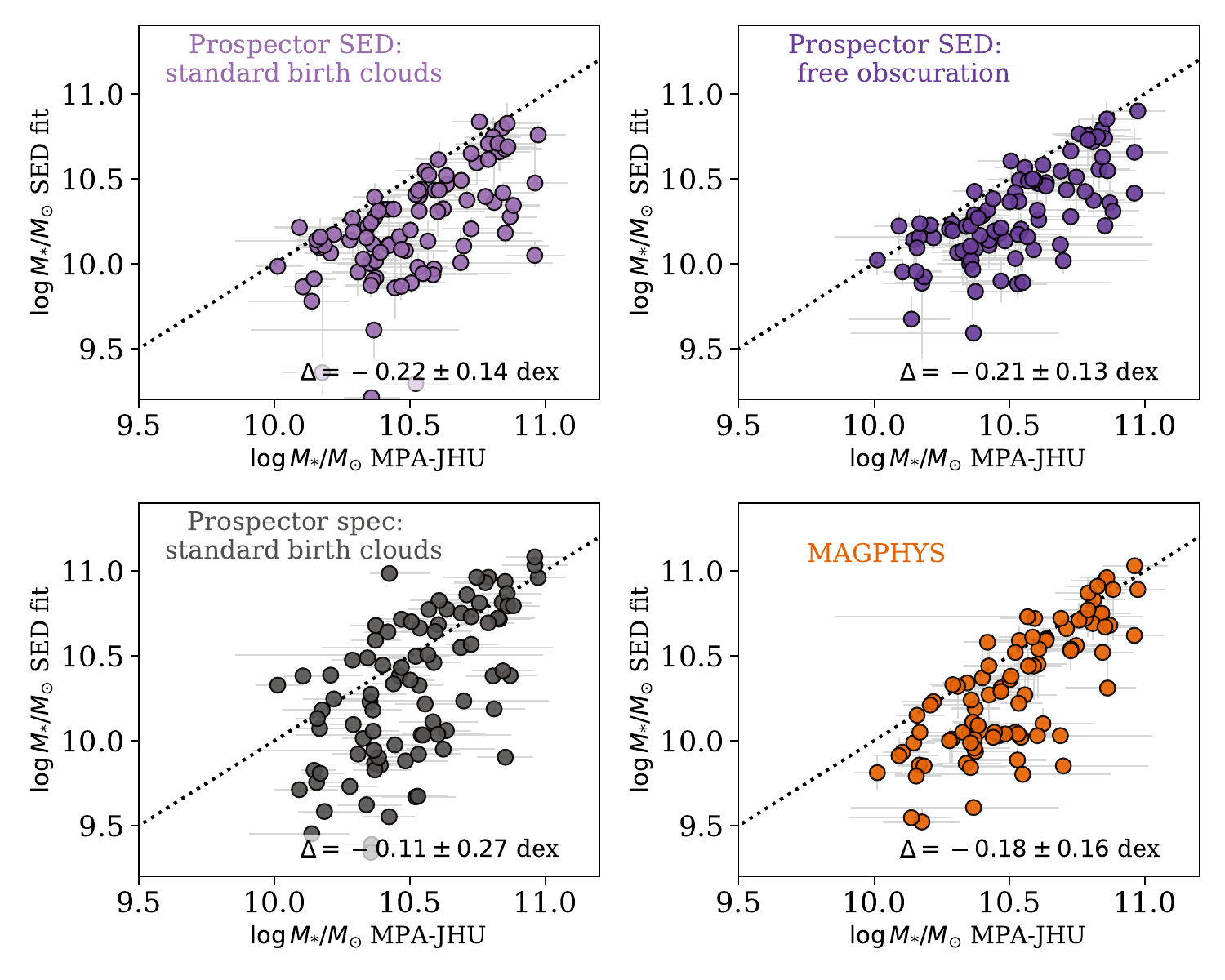}
\caption{\textbf{Comparison of stellar masses derived using different SPS fitting codes and modeling assumptions.} Each panel compares the best-fitting stellar masses from SPS fitting with the stellar mass in the SDSS value-added catalog (MPA-JHU; based on $ugriz$ photometry). Stellar masses obtained with \texttt{Prospector}+spec are scaled by the SDSS aperture correction. The fitting code, data used, and model assumptions are indicated at the top left of each panel. The black dashed line shows the 1:1 relation, and $\Delta$ gives the median and MAD of the logarithmic difference.}
\label{f:comparison_stellar_mass_SED_fits}
\end{figure*}

\subsection{Galaxy properties across different codes and assumptions}\label{sec:results:sed_fitting:different-codes}

In this section we compare the physical properties of the \textit{Sparks} galaxies derived under different codes and modeling assumptions (see details on model assumptions in section \ref{sec:SEDs:modeling}). We compare results from \texttt{MAGPHYS} and \texttt{Prospector} fits to the multi-wavelength SEDs, as well as \texttt{Prospector} fits to the rest-frame optical spectroscopy. For the \texttt{Prospector} SED fits we examine two dust attenuation models: the `standard birth-clouds' model, where young stars experience twice the attenuation of the older population, and the `free obscuration' model, where the attenuation of young stars is free to vary independently from the older population. The latter allows for the possibility of heavily obscured star formation that is invisible in the optical but emerges in the far-infrared. 

We compare our derived physical properties with those reported in the SDSS value-added catalogs (MPA-JHU; \citealt{b04, kauff03b, t04}). For stellar mass, we adopt the values from the MPA-JHU catalog, which were obtained by SPS modeling of the SDSS $ugriz$ photometry. While these masses are not expected to be more accurate than our own SED-based estimates, they provide a useful benchmark given their widespread use in the literature. We also make use of the MPA-JHU SFR estimates. For galaxies classified as star-forming based on optical line-ratio diagnostics, the SFR is derived from the reddening-corrected H$\alpha$ flux, further corrected for aperture losses. Because the H$\alpha$ emission is powered by ionizing radiation from O- and B-type stars, it is the most reliable tracer of ongoing star formation. For galaxies with composite, LINER, or Seyfert classifications, the MPA-JHU catalog instead estimates SFRs from the D$_n$4000 index; these estimates are much more uncertain, with a scatter of $\sim 0.4$ dex \citep{b04}. We therefore treat the H$\alpha$-based SFRs as our gold standard and use them as the primary benchmark for testing SFRs derived from our SPS and SED fitting.  By comparing against the H$\alpha$-based values in star-forming galaxies, we identify which approach yields the best agreement, and then adopt that method to estimate SFRs in galaxies where H$\alpha$ cannot be used (composites and AGN).

\begin{figure*}[t!]
	\centering
\includegraphics[width=0.7\textwidth]{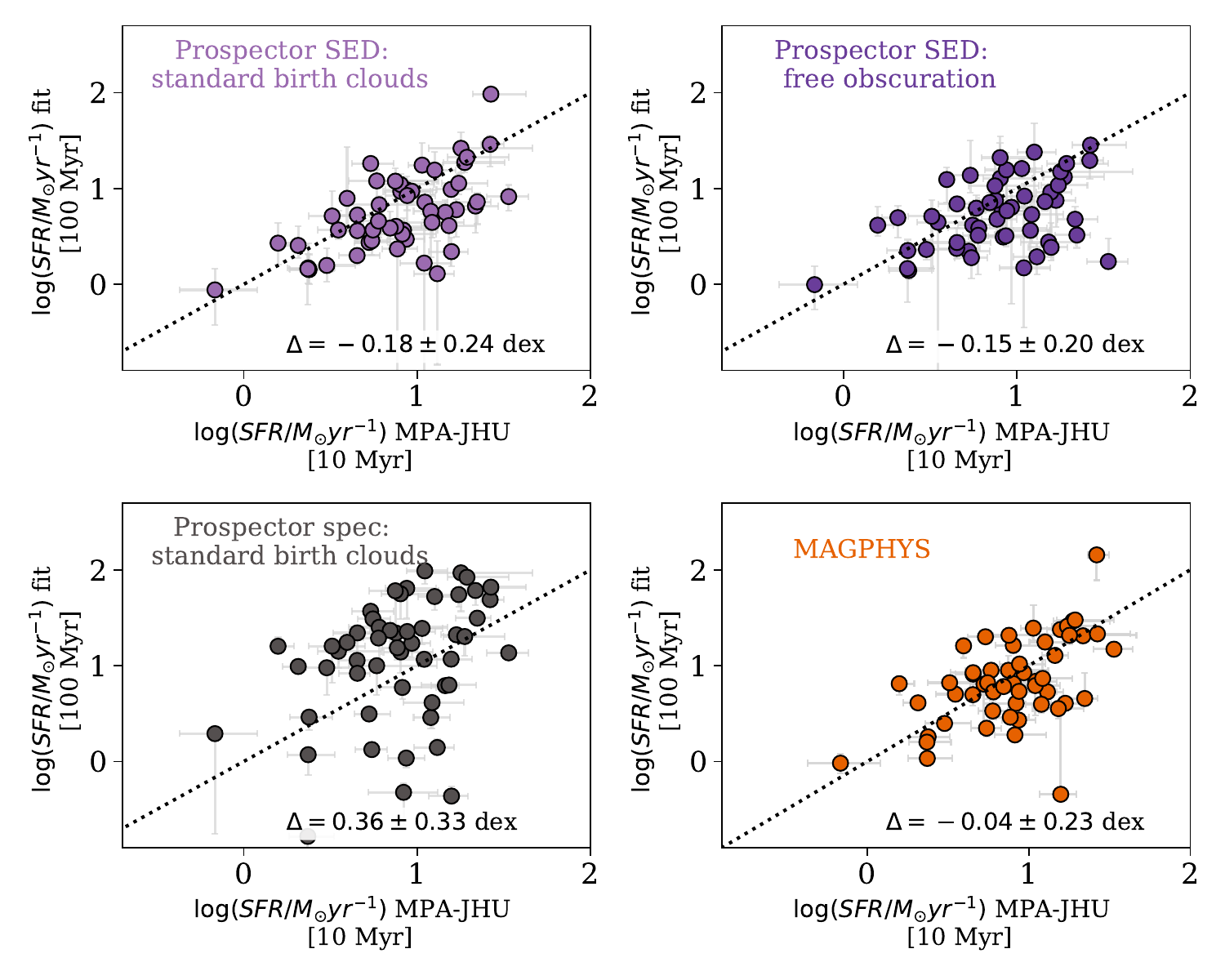}
\caption{\textbf{Comparison of SFR(100 Myr) derived using different SPS fitting codes and modeling assumptions for star-forming galaxies.} Each panel shows the best-fitting SFR, averaged over the past 100 Myr, obtained from SPS fitting versus the SDSS catalog SFR. The SFRs from \texttt{Prospector}+spec include the SDSS aperture correction. Only galaxies with line ratios consistent with star-formation ionization are included; the SDSS SFR is derived from the reddening-corrected H$\alpha$ flux, also corrected for aperture losses. H$\alpha$ traces the SFR averaged over the past 10 Myr. The black dashed line indicates the 1:1 relation, and $\Delta$ denotes the median and MAD of the logarithmic differences.}
\label{f:comparison_SFR_SED_fits}
\end{figure*}

We manually inspected the best-fitting spectra and SEDs and found that all four models provide reasonable fits. In \ref{app:sed_extra:sed_ir}, we compare the infrared luminosities predicted by the multi-wavelength SED models to those observed in the WISE W3, W4, and IRAS 60 \mic bands. Among the three SED fitting runs (\texttt{MAGPHYS} and two \texttt{Prospector} variants), \texttt{MAGPHYS} shows the best agreement with the observed mid- and far-infrared fluxes, with offsets $\leq 0.04$ dex and scatters $\leq 0.07$ dex. We attribute this to its greater flexibility in modeling the infrared regime--although we note that the infrared emission is likely under-constrained given the small number of data points relative to the many free parameters (see section \ref{sec:SEDs:modeling}). The two \texttt{Prospector} runs yield larger offsets of $[-0.09, 0.07]$ dex and scatters up to 0.17 dex. In particular, \texttt{Prospector} has difficulty reproducing the very red (W4-60 \mic) colors in some sources, overpredicting the W4 flux while underpredicting the 60 \mic flux. This is not entirely unexpected, since the \citet{draine07} dust emission models were built to describe typical star-forming galaxies rather than extreme starbursts and post-starbursts. Given that the overall offsets and scatters remain modest, we move forward with comparing the physical properties inferred by all three SED models.

In Figure \ref{f:comparison_stellar_mass_SED_fits} we compare stellar mass estimates from SPS and SED fitting to the MPA?JHU catalog values. The \texttt{Prospector}+spec masses are corrected using the SDSS aperture factors. All three SED-fitting approaches yield masses lower than the SDSS-derived values by 0.18?0.22 dex, with scatters of 0.13?0.16 dex. This offset is well established in the literature (e.g., \citealt{salim07, conroy13, leja19, lower20, tacchella22}) and is generally attributed to the simplistic SFHs assumed in $ugriz$-only fitting, which favor older populations and higher mass-to-light ratios, compared to the more flexible or non-parametric SFHs used in modern codes. The \texttt{Prospector}+spec fits show a smaller offset (0.11 dex) but larger scatter (0.27 dex), consistent with the fact that they rely only on optical data and lack infrared constraints. As shown in Section \ref{sec:results:sed_fitting:spec-vs-sed}, the increased scatter largely reflects differences in the best-fitting SFHs recovered from the spectroscopic versus photometric fits.

\begin{figure*}[t!]
	\centering
\includegraphics[width=0.7\textwidth]{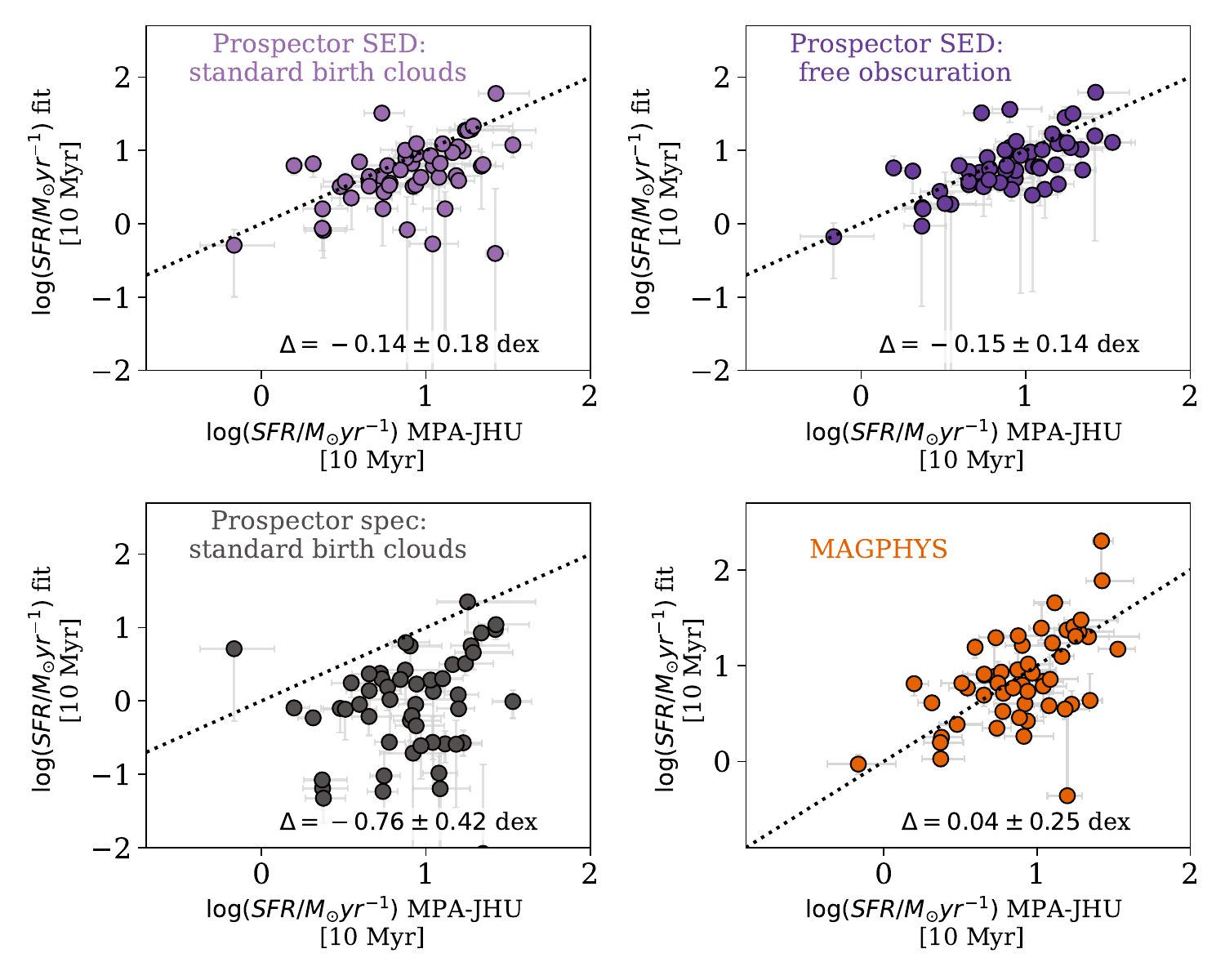}
\caption{\textbf{Comparison of SFR(10 Myr) derived using different SPS fitting codes and modeling assumptions for star-forming galaxies.} Each panel shows the best-fitting SFR, averaged over the past 10 Myr, obtained from SPS fitting versus the SDSS catalog SFR. SFRs from the spectroscopic fits include the SDSS aperture correction. Only galaxies with line ratios consistent with star-formation ionization are shown, where SFR is based on the H$\alpha$ line. The black dashed line indicates the 1:1 relation, and $\Delta$ gives the median and MAD of the logarithmic differences.}
\label{f:comparison_SFR_SED_fits_10Myr}
\end{figure*}

Next, we compare the SFRs derived from SPS fitting with the benchmark H$\alpha$-based SFRs from the SDSS, focusing on star-forming galaxies where H$\alpha$ emission is dominated by young, massive stars. Figure~\ref{f:comparison_SFR_SED_fits} presents the SFRs averaged over the past 100 Myr for the different fitting approaches. Because these galaxies are on the starburst-to-post-starburst sequence, SFR(100 Myr) may not necessarily match the instantaneous H$\alpha$-based SFR. Instead, deviations can reveal recent SFH variations. Across all three SED-fitting methods, the offset between SFR(100 Myr) and SFR(10 Myr) shows median offsets of -0.18 to -0.4 dex, with a scatter of 0.20?0.24 dex. In contrast, SFR(100 Myr) from the \texttt{Prospector} fits to the optical spectra is systematically higher, with a larger scatter of 0.33 dex. As discussed in Section~\ref{sec:results:sed_fitting:spec-vs-sed}, strong Balmer absorption in the spectra drives the fits toward declining SFHs, producing elevated SFRs averaged over 100 Myr but much lower SFRs over the recent 10 Myr.

In figure~\ref{f:Pros_comparison_SFRs_different_scales_vs_SFH} (\ref{app:sed_extra:sfr100_vs_sfr10}), we investigate whether the scatter between SFR(100 Myr) and SFR(10 Myr) is related to a galaxy?s evolutionary stage by color-coding the points according to starburst age and offset from the star-forming main sequence. For two of the three \texttt{Prospector} models (photometric and spectroscopic fits with standard birth clouds), we find significant negative correlations between $\Delta y = \log(\mathrm{SFR}{100,\mathrm{Myr}}) - \log(\mathrm{SFR}{10,\mathrm{Myr}})$ and $t_{\mathrm{SB}}$, indicating that SFR(100 Myr) tends to fall below SFR(10 Myr) for older starburst ages. This trend would be expected in the case that the post-burst population becomes increasingly older than 100 Myr. Despite the significance of these correlations, we caution against over-interpreting this trend because (i) multi-wavelength SEDs are relatively insensitive to detailed SFHs over the past $\sim$1 Gyr, making $t_{\mathrm{SB}}$ estimates from photometry uncertain, and (ii) SFHs derived from optical spectroscopy are also limited, as they are shown to underestimate the instantaneous SFR, while boosting the SFR over the past 100 Myrs, which might impact the estimated $t_{\mathrm{SB}}$ to some extent (see section \ref{sec:results:sed_fitting:spec-vs-sed}).

\begin{figure*}
	\centering
\includegraphics[width=1\textwidth]{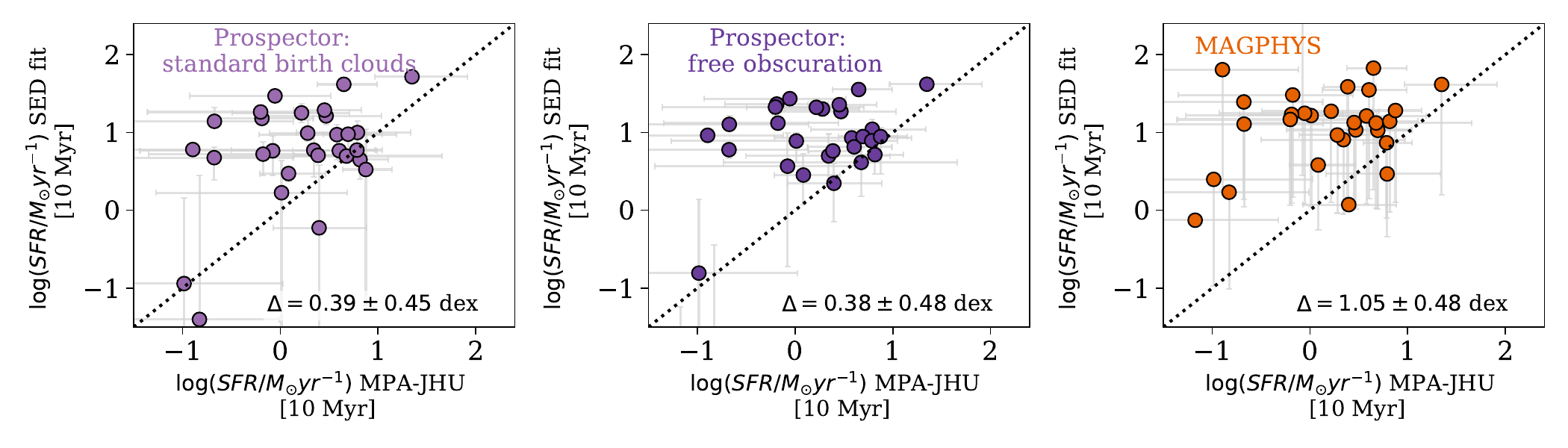}
\caption{\textbf{Multi-wavelength SED fitting-based SFR(10 Myr) for \textit{non} star-forming galaxies.} The panels show SFR(10 Myr) derived from the best-fitting SED models versus the SDSS catalog SFR. \texttt{Prospector}+spec results are not shown as we believe these do not trace the ongoing star formation accurately. Only galaxies with line ratios inconsistent with star formation are included (composites, AGN, or weak-line galaxies); the SDSS SFR is based on the D$_n$4000\AA, index and is highly uncertain. The black dashed line marks the 1:1 relation, and $\Delta$ indicates the median and MAD of the logarithmic differences.}
\label{f:comparison_SFR_SED_fits_composites_and_AGN_10Myr}
\end{figure*}

Having examined SFR(100 Myr) from SPS fitting, we now turn to SFR(10 Myr). In figure~\ref{f:comparison_SFR_SED_fits_10Myr}, we compare the SFRs predicted by the different models with those derived from the H$\alpha$ flux for star-forming galaxies. All the panchromatic SED fitting models perform reasonably well. The \texttt{Prospector} free obscuration model shows a median offset of $-0.15$ dex and a scatter of $0.14$ dex, only slightly better than the standard birth-clouds model. This suggests that for most galaxies in the sample, additional obscured star formation beyond what can be modeled under the standard birth-clouds assumption is not required. The \texttt{MAGPHYS} fits result in a smaller median offset of $0.04$ dex, but a larger scatter of $0.25$ dex around the one-to-one line.

In contrast, the \texttt{Prospector}+spec model, which fits the optical continuum emission, underpredicts the instantaneous SFR, with a median offset of $-0.76$ dex and a scatter of $0.42$ dex\footnote{As described in Section~\ref{sec:SEDs:prospector}, the fits do not tie the \halpha flux to the stellar continuum emission as 40 of the 93 \textit{Sparks} galaxies show optical line ratios inconsistent with pure star formation.}. Combined with the results from Section~\ref{sec:results:sed_fitting:spec-vs-sed}, which show that it underpredicts the observed far-infrared emission, we conclude that SPS fitting of the optical continuum alone, without additional multi-wavelength tracers or self-consistent modeling of the \halpha flux, does not reliably recover the ongoing star formation in the \textit{Sparks} galaxies, likely because the fit is driven by the strong Balmer features to prefer intermediate-age stellar populations at the expense of ongoing star formation.

We therefore adopt \texttt{Prospector}+SED to estimate the instantaneous SFRs in galaxies with line ratios inconsistent with pure star formation (composites, AGN, and weak-lined systems). Measuring their SFRs is particularly important for understanding galaxy?black hole coevolution and the role of AGN feedback. As discussed earlier, the SDSS estimates based on the D$_{n}$4000\AA\, index are highly uncertain, motivating the use of alternative approaches. In figure~\ref{f:comparison_SFR_SED_fits_composites_and_AGN_10Myr} we present SFR(10 Myr) for these galaxies, and for completeness, figure~\ref{f:comparison_SFR_SED_fits_composites_and_AGN} in Appendix~\ref{app:sed_extra:sfr100_vs_sfr10} shows the corresponding SFR(100 Myr). Across different SPS fitting codes and assumptions, we find that multi-wavelength SEDs predict instantaneous SFRs that are systematically higher by 0.3?0.7 dex compared to the D$_{n}$4000\AA-based estimates. This is a lower limit as we showed that there is an offset of $-0.15$ dex between SFR(10 Myr; SED) and the H$\alpha$-based SFR in star-forming galaxies. For the composite galaxies, we also derive SFR(10 Myr) using the H$\alpha$ decomposition method of \citet{wild10}, which separates star-formation and AGN contributions along the [NII]/H$\alpha$ BPT mixing sequence\footnote{This technique assumes that galaxies in the composite region lie along a mixing sequence between star formation and AGN ionization. A galaxy?s location along this sequence is used to estimate the fraction of H$\alpha$ emission powered by star formation, yielding a corrected H$\alpha$-based SFR.}, and find reasonable agreement given the uncertainties. In Paper I, we use these new SFR estimates to update the position of \textit{Sparks} galaxies with BPT classification other than HII with respect to the star-forming main sequence.

\begin{figure}
	\centering
\includegraphics[width=\columnwidth]{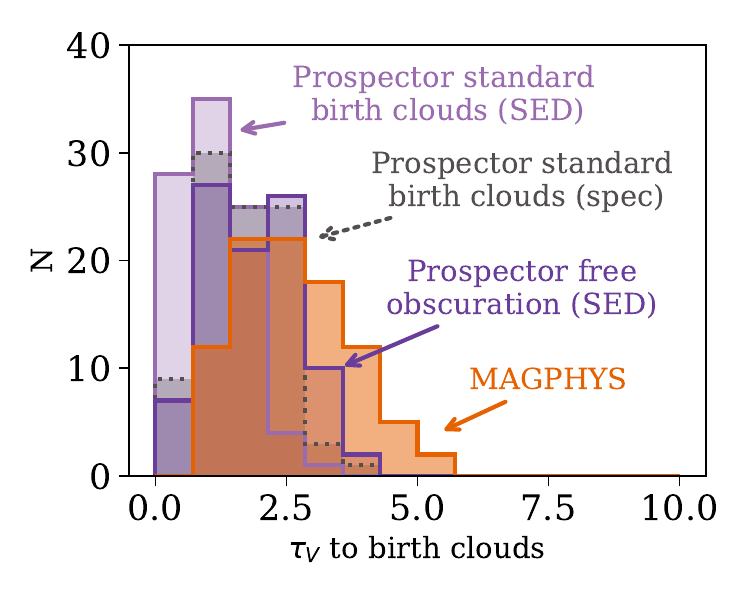}
\caption{\textbf{Dust optical depth toward birth clouds across SPS codes and modeling assumptions.} The histograms show the distribution of $\tau_{V}$?the optical depth toward young stellar populations?derived from the different SPS fitting codes indicated in the figure.}
\label{f:tau_comparison_SED_fits}
\end{figure}

Figure \ref{f:tau_comparison_SED_fits} compares the dust optical depth toward the young stellar populations ($< 10$ Myr) derived using the different codes and assumptions. The figure shows that different codes, input datasets, and modeling assumptions, have an impact on the derived dust reddening in transitioning sources. Ordered from lowest to highest median $\tau_{V}$: \texttt{Prospector}+SED standard birth-clouds with 1.16 mag; \texttt{Prospector}+spec standard birth-clouds with 1.73 mag; \texttt{Prospector}+SED free obscuration with 1.93 mag; and \texttt{MAGPHYS} with 2.55 mag. The differences between \texttt{Prospector} and \texttt{MAGPHYS} probably reflect the differences in assumed SFH forms, and priors on dust attenuation and infrared emission. Specifically, \texttt{MAGPHYS} models the SED using younger stellar populations and higher SFRs, obscured by larger columns of dust. This is in line with a recent study that compared the output of different SED fitting codes, finding an age-dust degeneracy across codes \citep{pacifici23}.

\begin{figure*}
	\centering
\includegraphics[width=0.9\textwidth]{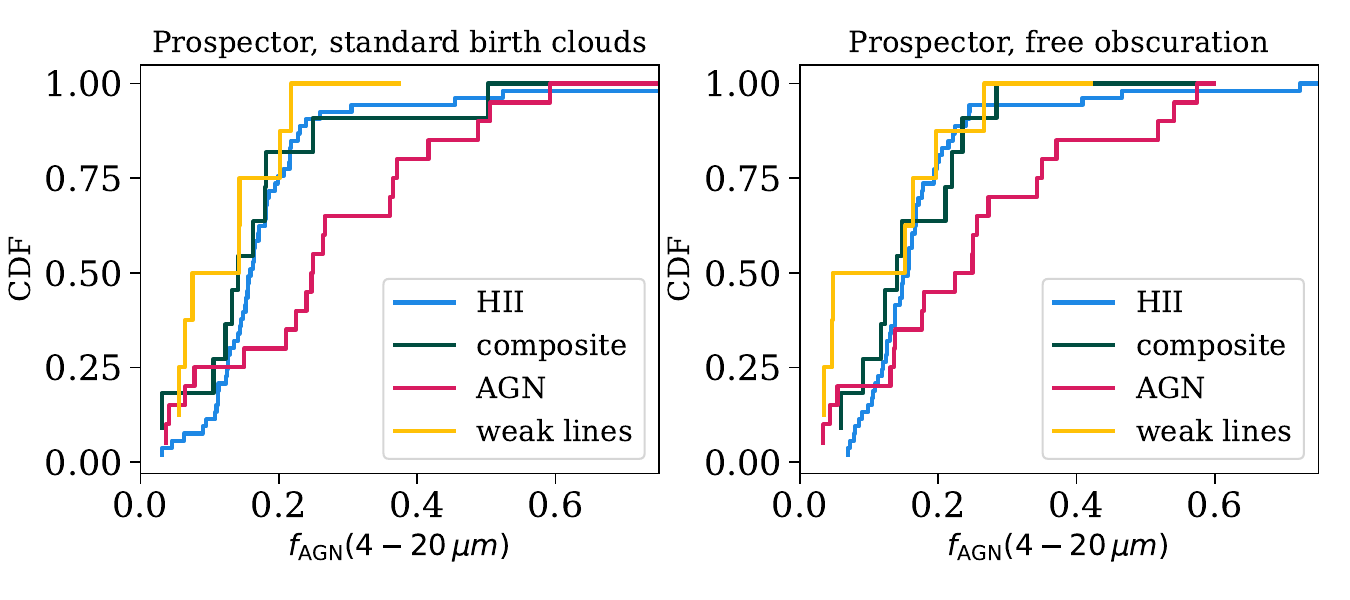}
\caption{\textbf{Cumulative distribution function (CDF) of the AGN torus contribution to mid-infrared wavelengths.} The panels compare the AGN torus contribution in the $4-20$ \mic\xspace wavelength range, for galaxies with different emission line ratio classifications (star-forming; composite; AGN; and weak lines). The left panel shows the results from the standard birth-clouds model, and the right from the free obscuration model.}
\label{f:torus_to_total_W4_CDFs}
\end{figure*}

\subsection{Characterization of the torus component in \texttt{Prospector}}\label{sec:results:sed_fitting:agn-torus}

One advantage of \texttt{Prospector} over \texttt{MAGPHYS} in modeling galaxy SEDs is its ability to include an AGN torus component at mid-infrared wavelengths. Because torus emission overlaps with dust heated by young, massive stars, neglecting the AGN contribution can bias estimates of the obscured star formation rate (e.g., \citealt{rosario12, lusso13, ciesla15, kirkpatrick15, lani17}). This is particularly important for post-starburst galaxies: in some samples, significant infrared emission has been observed and linked to residual star formation (\citealt{baron22, baron23}), while in others it is attributed primarily to AGN activity (\citealt{alatalo16b, wu23}). Moreover, the torus component can help identify obscured accreting black holes, potentially tracing a different stage in the evolution of starburst mergers (e.g., \citealt{spoon07, armus09, imanishi09, goulding09, mateos13} and review by \citealt{hickox18}).

The main limitation of the \texttt{Prospector} torus model is the absence of a corresponding ultraviolet?optical accretion disk component. Since the torus mid-infrared emission is powered by ultraviolet?optical photons from the accretion disk that are absorbed and re-emitted by dust (e.g., \citealt{netzer15, hickox18}), modeling only the torus breaks the energy-balance constraint: the absorbed ultraviolet?optical luminosity no longer matches the infrared luminosity. Consequently, the model may misattribute residual mid-infrared emission to an AGN torus rather than to dust heated by obscured star formation, potentially biasing inferred SFRs. In the absence of energy-balance constraints, it is therefore essential to benchmark the \texttt{Prospector} torus component against independent diagnostics of black hole accretion, and to compare the derived torus properties with those inferred for mid-infrared?selected AGN.


In this section, we examine the contribution of torus infrared emission to the best-fitting \texttt{Prospector} SEDs and compare it with optical line ratio diagnostics, which trace AGN activity through hard ionizing radiation. While a one-to-one correspondence between AGN emission-line ratios and torus contributions to the mid-infrared is not expected (e.g., \citealt{hickox09, hickox18, hviding22}), we expect some degree of correspondence, as both optical lines and mid-infrared emission trace the dominant source of radiation field. The \textit{Sparks} galaxy sample contains no type I AGN, as it was selected to include stellar-continuum?dominated sources with strong Balmer absorption lines. The sample does include 21 type II AGN, whose optical line ratios are consistent with Seyfert emission, and 11 composite galaxies, whose line ratios indicate a combination of star formation and AGN ionization.

In Figure~\ref{f:torus_to_total_W4_CDFs} we show the cumulative distribution functions (CDFs) of the torus contribution to the mid-infrared emission by integrating within the wavelength range $4-20$ \mic as in \citet{leja18}. The galaxies are divided according to their optical line-ratio classifications. A steeply rising CDF at low AGN-to-total fractions indicates that most sources in that class have only a small torus contribution, whereas a more gradual rise indicates larger AGN fractions. For both \texttt{Prospector}+SED models and across all galaxy classes, the CDFs of the purely star-forming, composite, and weak-line galaxies reach \(>0.75\) by an AGN-to-total contribution of \(\leq 0.2\), implying that most of them do not require a strong AGN component in the infrared. The CDFs of the AGN hosts rise more gradually, suggesting higher AGN-to-total contributions. 

We quantified the trends in Figure~\ref{f:torus_to_total_W4_CDFs} using the Kolmogorov--Smirnov (KS; \citealt{press07}) statistic to test the null hypothesis that two distributions are drawn from the same parent population. We find that the purely star-forming galaxies and composites are statistically indistinguishable, with KS statistic 0.160 and $p$-value 0.9402. For the star-forming and the AGN hosts, we are able to reject the null hypothesis, with KS statistic 0.537 and $p$-value 0.0002. We therefore conclude that \texttt{Prospector} identifies larger torus contributions, on average, for sources with optical line ratios consistent with AGN. 

\begin{figure*}
	\centering
\includegraphics[width=1\textwidth]{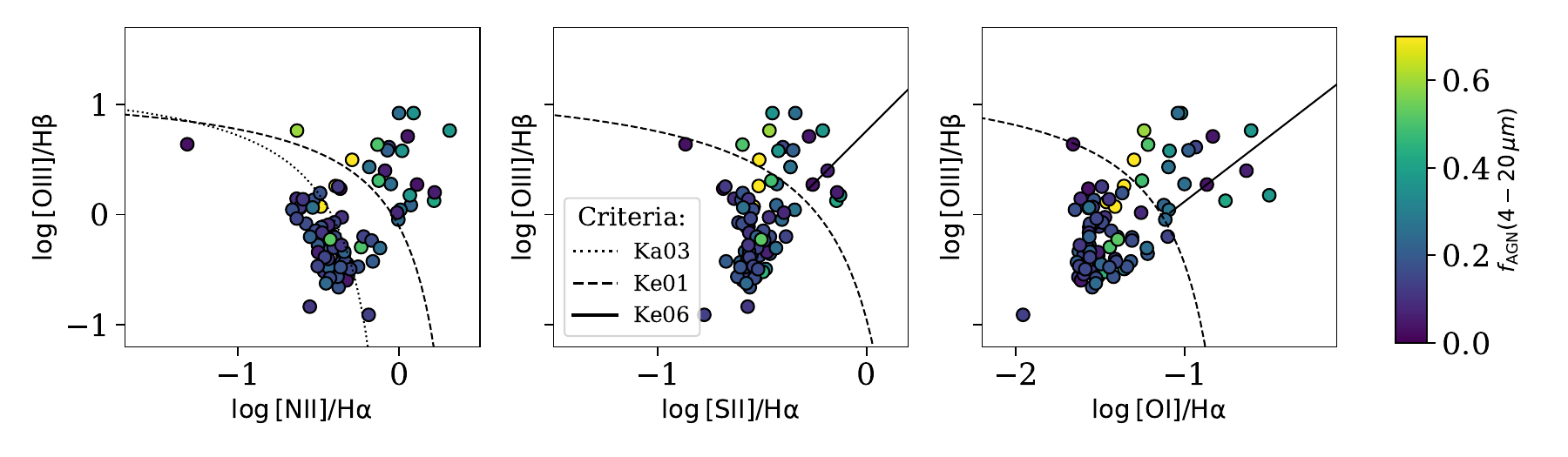}
\caption{\textbf{Optical line ratio diagnostic diagrams color-coded by the torus mid-infrared contribution.} The panels show the standard BPT line ratios, \oiiihbeta{} versus \niihalpha{}, \siihalpha{}, and \oihalpha{}, for the 93 galaxies in our sample, using emission-line measurements from the MPA-JHU catalog \citep{kauff03b, b04, t04}. The points are color-coded according to the torus-to-total contribution within the $4-20$ \mic wavelength range. The dividing curves commonly used to separate star-forming and AGN-dominated galaxies from \citet{kewley01} (Ke01) and \citet{kauff03a} (Ka03), as well as the LINER?Seyfert demarcation from \citet[Ke06]{kewley06}, are shown for reference.}
\label{f:BPT_diagram_ftorus}
\end{figure*}

In Figure~\ref{f:BPT_diagram_ftorus} we show the Sparks galaxies on optical line diagnostic diagrams, color-coded by the torus contribution to the mid-infrared. Interestingly, \texttt{Prospector} assigns some mid-infrared fraction to torus emission in purely star-forming, composite, and weak-line galaxies. There are 5 (10\%), 2 (18\%), and 1 (12\%) galaxies with a torus contribution larger than 30\% for the star-forming, composite, and weak-line galaxies, respectively. One approach to test the significance of this mid-infrared emission is to fit these galaxies both with and without a torus component and compare the resulting SEDs. In addition, the FIRE near-infrared spectra collected for the \textit{Sparks} survey provide an opportunity to search for signatures of obscured AGN missed in the optical, such as coronal lines tracing highly ionized gas, very broad Paschen-line components from the broad-line region, or [Fe II] and H$_2$ emission properties characteristic of AGN activity (e.g., \citealt{riffel13}). We intend to explore these specific sources in a future publication on the AGN properties in the \textit{Sparks} galaxies.

\begin{figure*}
	\centering
\includegraphics[width=1\textwidth]{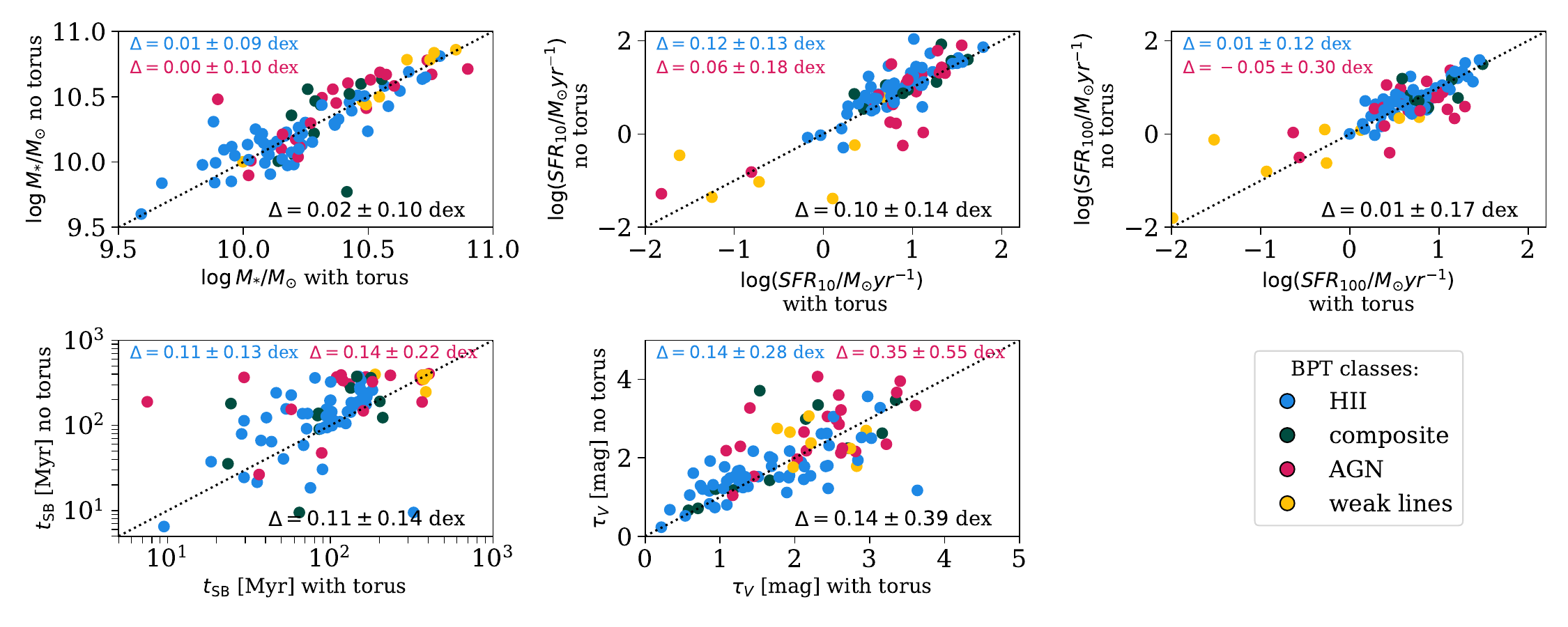}
\caption{\textbf{Comparison of galaxy physical properties derived with and without the AGN torus component.} The panels show the galaxy properties derived using the \texttt{Prospector} standard birth-clouds model with and without the AGN torus component. The points are color-coded according to the MPA-JHU classification of the source. The black dashed line marks the 1:1 relation, and $\Delta$ gives the median and MAD of the logarithmic difference: black (bottom) for the entire sample, and blue and magenta (top) for the star-forming and AGN hosts respectively.}
\label{f:comparison_of_derived_properties_with_and_without_torus}
\end{figure*}

Next, we compare the galaxy physical properties derived with and without an AGN torus component. To do so, we ran the \texttt{Prospector} standard birth-clouds model with the AGN torus component turned off. In Figure~\ref{f:comparison_of_derived_properties_with_and_without_torus}, we compare the resulting stellar masses, SFRs, starburst ages, and dust optical depths between the two runs. For the stellar mass, we find a median offset of 0.02 dex and a scatter of 0.10 dex, with comparable values for both the MPA?JHU classified star-forming galaxies and AGN hosts. This indicates that stellar masses are recovered robustly regardless of whether an AGN torus component is included. For the SFR averaged over the past 10 Myr, we find a median offset of 0.10 dex and a scatter of 0.14 dex, while for the SFR averaged over the past 100 Myr we find an offset of 0.01 dex and a scatter of 0.17 dex. In both cases, the scatter is larger for AGN hosts than for the full sample. Additionally, when the AGN component is turned off, the SFR averaged over 10 Myr shows a median offset of 0.12 dex for the purely star-forming galaxies relative to the run that includes the AGN component.

For the starburst age, we find a median offset of 0.11 dex and a scatter of 0.14 dex, with AGN hosts exhibiting a larger scatter (0.22 dex) than the full sample. Among all the examined physical parameters, the largest differences are found for the derived optical depth toward birth clouds, with a median offset of 0.14 dex and a scatter of 0.39 dex for the full sample. Interestingly, both the offset and scatter are smaller for star-forming galaxies ($0.14 \pm 0.28$ dex) than for AGN hosts ($0.35 \pm 0.55$ dex). This behavior is consistent with the expectation that, in the absence of an explicit AGN mid-infrared component, the model compensates by increasing the optical depth toward young stellar populations in order to reproduce the observed mid-infrared emission. As a result, the inferred optical depths can differ substantially in systems with strong mid-infrared emission, as is often the case for AGN hosts. 

\begin{figure*}
	\centering
\includegraphics[width=0.9\textwidth]{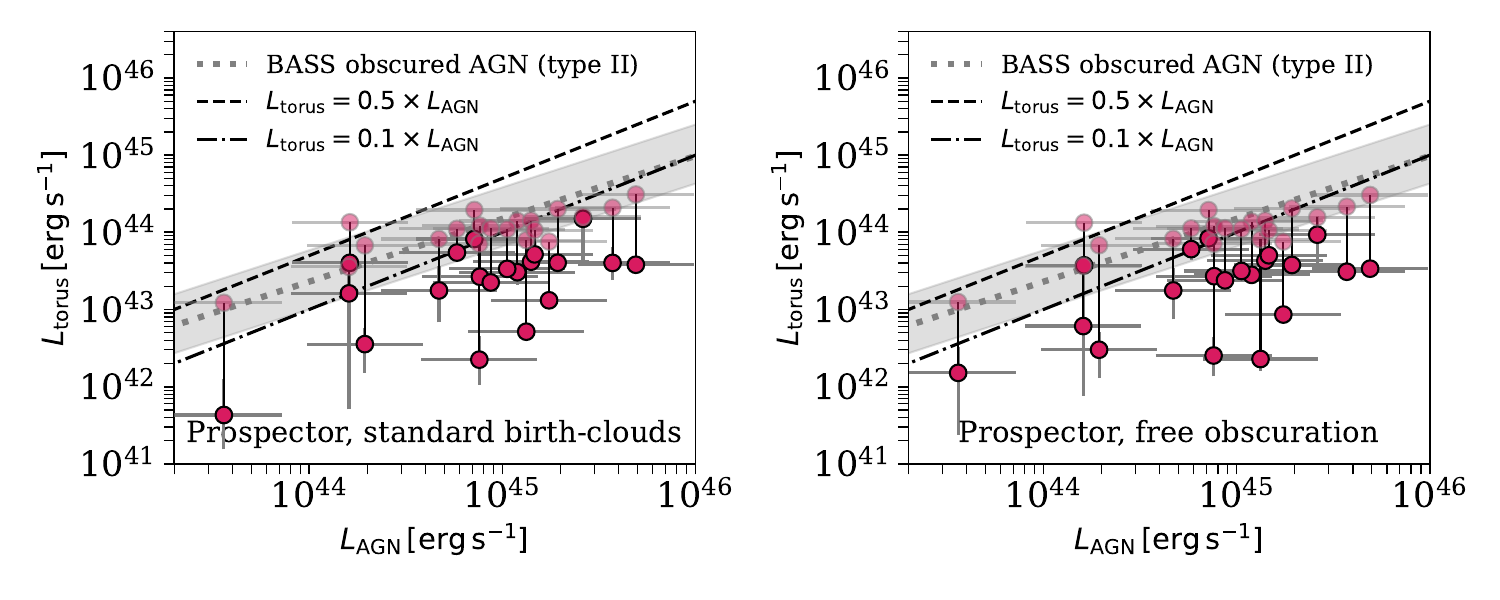}
\caption{\textbf{Predicted torus luminosity versus AGN bolometric luminosity.} Dark magenta points show the total infrared torus luminosity, derived from the best-fitting \texttt{Prospector} standard birth-clouds (left) and free-obscuration (right) models, plotted against the AGN bolometric luminosity estimated from optical emission lines. The lighter magenta points, connected to the darker points by vertical lines, indicate the maximum torus luminosity allowed by the mid-infrared observations under the assumption that the torus completely dominates the mid-infrared. Only AGN hosts are shown. The black dashed and dash-dotted lines mark $L_{\mathrm{torus}} = 0.5 \times L_{\mathrm{AGN}}$ and $L_{\mathrm{torus}} = 0.1 \times L_{\mathrm{AGN}}$, respectively. The best-fitting linear relation between $\log L_{\mathrm{torus}}$ and $\log L_{\mathrm{AGN}}$ obtained for the BASS type~II AGN is shown with a gray dotted line, with the 16th-84th percentiles shown as a light gray band. The \texttt{Prospector}-derived $L_{\mathrm{torus}}$ values are about an order of magnitude lower given their $\log L_{\mathrm{AGN}}$ compared to other samples.}
\label{f:Ltorus_vs_LAGN}
\end{figure*}

Next, we characterize the properties of torus emission in optically selected AGN hosts and compare them to those observed in other AGN samples. We do not include composites in this analysis. In Figure \ref{f:Ltorus_vs_LAGN}, we examine the relation between the torus infrared luminosity and the AGN bolometric luminosity. Previous studies of AGN infrared properties typically find $L_{\mathrm{torus}}/L_{\mathrm{AGN}}$ in the range $\sim 0.1-0.5$, including type~I and type~II AGN (e.g., \citealt{mullaney11, mor12, lusso13, netzer16, stalevski16, baron19a}). Within the BASS survey (BAT AGN Spectroscopic Survey; \citealt{koss17, koss22}), which provides a census of the brightest hard-X-ray-selected AGNs in the local Universe, the median (16th, 84th percentiles) $L_{\mathrm{torus}}/L_{\mathrm{AGN}}$ is 0.24 (0.10, 0.61) for unobscured type I AGN and 0.18 (0.07, 0.51) for obscured type II AGN \citep{ricci23}. In Figure~\ref{f:Ltorus_vs_LAGN} we show the best-fitting linear relation between $\log L_{\mathrm{torus}}$ and $\log L_{\mathrm{AGN}}$ derived for the BASS type~II AGN\footnote{Here we define obscured type~II AGN as systems with obscuring column $\log N_{\mathrm{H}}/\mathrm{cm^{-2}} > 22$ (see \citealt{ricci23} for details).}, marking the 16th to 84th percentiles with a light gray band. 

We derive the torus luminosities by integrating the best-fitting \texttt{Prospector} torus models, while estimating the AGN bolometric luminosity from the dust-corrected \hbeta line using the bolometric correction factor from \citet{netzer19}\footnote{We examined other bolometric correction factors based on \oiii and \oi, and found consistent luminosities \citep{netzer09, heckman14}.}. The figure shows that, for a given AGN bolometric luminosity, the \texttt{Prospector} torus luminosity is about an order of magnitude lower than that observed in other samples. The median (16th, 84th percentiles) $L_{\mathrm{torus}}/L_{\mathrm{AGN}}$ derived with \texttt{Prospector} is 0.03 (0.007, 0.09).

What drives the unusually low $L_{\mathrm{torus}}/L_{\mathrm{AGN}}$ values derived using \texttt{Prospector}? We first quantify the maximum $L_{\mathrm{torus}}$ allowed by our observations as independently as possible from the \texttt{Prospector} torus model assumptions. Assuming that the torus completely dominates the mid-infrared, we can integrate the best-fitting total emission within $4-20$ \mic and estimate $L_{\mathrm{torus}}$ from that. The \texttt{Prospector} torus models indicate $L_{\mathrm{torus}} = 1.7 \times L_{\mathrm{torus}}(4-20\, \mu m)$, with the exact factor ranging from 1.5 to 2 depending on the torus optical depth (also in line with the \citealt{stalevski12} torus models). This range is also consistent with observations of type~I AGN, where the mid-infrared wavelength range is dominated by torus emission (e.g., \citealt{mor12, lani17}). The maximum allowable torus luminosities are shown as light magenta points in Figure~\ref{f:Ltorus_vs_LAGN}. Even under the extreme assumption that the torus dominates the mid-infrared, the derived $L_{\mathrm{torus}}/L_{\mathrm{AGN}}$ remains low, with a median (16th, 84th percentiles) of 0.11 (0.06, 0.27). This suggests that the observed WISE mid-infrared luminosities cannot accommodate $L_{\mathrm{torus}}/L_{\mathrm{AGN}}$ much larger than $\sim 0.1$.

The unusually low $L_{\mathrm{torus}}/L_{\mathrm{AGN}}$ values may instead reflect reduced torus covering factors in the \textit{Sparks} AGN hosts, potentially linked to their evolutionary stage along the starburst-to-post-starburst sequence. The torus infrared luminosity represents the fraction of accretion disk emission that is reprocessed by dust in the torus. Thus, $L_{\mathrm{torus}}/L_{\mathrm{AGN}}$ is expected to trace the fraction of the sky obscured by dust?the torus covering factor (e.g., \citealt{mor12, netzer16, stalevski16, lani17, ricci23}). In clumpy two-phase torus models, the exact relation between $L_{\mathrm{torus}}/L_{\mathrm{AGN}}$ and the covering factor depends on the specific torus parameters \citep{stalevski16}. For type~II AGN, our observed $L_{\mathrm{torus}}/L_{\mathrm{AGN}} \sim 0.03$ corresponds to covering factors of $0.28$?$0.34$, depending on the mid-infrared optical depth. These values are lower than the $\sim 0.6$?$0.7$ covering factors inferred for the general AGN population \citep{stalevski16} and those of obscured ($\log N_{\mathrm{H}}/\mathrm{cm^{-2}} > 22$) type~II AGN observed in the BASS survey \citep{ricci23}.

Statistical studies of X-ray?selected AGN show that the torus covering factor decreases in systems with high black hole accretion rates, consistent with radiation pressure expelling the Compton-thin regions of the torus and leaving behind only Compton-thick material with a smaller covering factor (e.g., \citealt{ricci17, ricci23} and references therein). At the same time, along the merger sequence, the fraction of Compton-thick AGN reaches its maximum near galaxy coalescence \citep{ricci21}.

Adopting the new estimates of SFR and SFH derived in this work, in Paper~I we showed that the \textit{Sparks} AGN hosts are consistent with galaxies undergoing a second starburst following a quenching of an earlier starburst taking place a few hundreds of Myrs ago. Their images provide first hints that their hosts may be systems close to the coalescence stage of a merger. The low $L_{\mathrm{torus}}/L_{\mathrm{AGN}}$ ratios observed in our sources may indicate that these systems are already past the blow-out phase, where only Compton-thick material with a small covering factor remains (e.g., \citealt{ricci17, blecha18, ricci21, kawaguchi20}). Confirming this scenario requires X-ray observations to constrain the column density of the obscuring material along the line of sight to the AGN. Such observations would also provide an independent test of the reliability of our AGN bolometric luminosity estimates, which are currently based on narrow emission lines.

\pagebreak

\section{Summary}\label{sec:summary}

Deriving the physical properties of galaxies, such as their stellar masses, SFRs, and detailed SFHs, requires modeling their integrated light. This is typically achieved through stellar population synthesis fitting of spectra or through multi-wavelength SED fitting, which incorporates the galaxy?s full energy budget from the far-ultraviolet to the far-infrared. Several codes exist for this purpose, including \texttt{MAGPHYS}, \texttt{Prospector}, \texttt{BAGPIPES}, and \texttt{CIGALE}, each employing different assumptions and modeling ingredients for stellar evolution, SFHs, dust attenuation, and AGN emission. Given these differences, it is crucial to benchmark these codes against one another and on diverse galaxy samples to understand which physical properties can be recovered robustly and to quantify potential systematic biases.

This paper studies the properties of galaxies from the \textit{Sparks} survey, a sample of 93 local massive galaxies selected to span the starburst-to-post-starburst evolutionary sequence (Baron et al. submitted). Here we collect panchromatic photometry from far-ultraviolet to far-infrared and fit the multi-wavelength SEDs and optical spectra using the codes \texttt{Prospector} and \texttt{MAGPHYS} with varying model assumptions. Our goals are twofold: (I) to assess the robustness of derived parameters, such as stellar mass, SFR, and SFH parameters, for galaxies in this extreme and rapidly evolving phase, where fitting choices are particularly consequential; and (II) to identify the set of physical parameters we will adopt as the final properties for the \textit{Sparks} survey. Our results are summarized below.

 \textbf{1. Fitting optical spectra versus multi-wavelength photometry with \texttt{Prospector} reveals systematic differences in derived galaxy properties, driven by different best-fitting star formation histories (section~\ref{sec:results:sed_fitting:spec-vs-sed}).} We compare the physical properties derived from fitting rest-frame optical spectra versus far-ultraviolet to far-infrared photometry using \texttt{Prospector}, with similar model ingredients and priors. While the stellar mass is relatively comparable (median offset of $0.1$ dex and scatter of $0.24$ dex), we find larger differences in derived SFRs. The spectroscopic fits yield SFRs with median offset of $+0.3$ dex ($0.4$ dex scatter) over 100 Myr, and $-0.8$ dex ($0.5$ dex scatter) over 10 Myr, compared to photometric fits. The  offsets and scatter are not driven by aperture mismatch but by differences in the best-fitting SFHs. Fits to the optical continuum emission underpredict the far-infrared luminosity (median offset $-0.6$ dex), demonstrating that it misses ongoing star formation. These trends are consistent with expectations given the strong Balmer absorption features in the optical spectra, which drive the fits toward rapidly declining SFHs at the expense of ongoing star-formation, and suggest that infrared constraints may be helpful for reliably recovering the full SFHs of galaxies transitioning from starburst to post-starburst.

 \textbf{2. Comparing physical properties across different codes and assumptions reveals comparable stellar masses but model-dependent SFRs and dust content (section~\ref{sec:results:sed_fitting:different-codes}).} We benchmark \texttt{MAGPHYS} and \texttt{Prospector} by comparing their outputs with each other and with properties derived from the SDSS. Stellar masses are relatively comparable across the different SED-fitting models, all yielding values lower than the MPA-JHU catalog by $\sim 0.2$ dex, a known offset attributed to more flexible SFHs. We find larger discrepancies in SFRs and dust attenuation; \texttt{MAGPHYS} generally infers higher SFRs and more obscuration than \texttt{Prospector}, and while it provides a better fit to the infrared photometry, we suspect its flexible infrared modeling may be under-constrained by our data. To identify the most reliable approach to estimate the instantaneous SFR, we compare the derived SFRs against \halpha-based estimates for the star-forming galaxies in our sample. We find that the instantaneous SFR (averaged over 10 Myr) from \texttt{Prospector}+SED fitting provides a reasonable agreement, with a median offset of $-0.15$ dex and a scatter of $0.14$ dex. We therefore adopt these SFRs for galaxies whose line ratios are inconsistent with pure star formation, finding that composite and AGN galaxies have systematically higher star formation activity than implied by their uncertain D$_n$4000\AA-based catalog estimates.

To summarize, for the final properties of the  \textit{Sparks} galaxies, we adopt \halpha-based SFRs for the star-forming galaxies and the \texttt{Prospector}+SED SFRs for galaxies whose line ratios are inconsistent with pure star-formation. We further adopt the \texttt{Prospector}+SED-based stellar masses, SFH parameters such as $t_{\mathrm{SB}}$, and dust reddening, for all of \textit{Sparks} galaxies.

\textbf{3. The \texttt{Prospector} AGN torus model distinguishes optically-classified AGN but infers unusually low infrared luminosities, potentially reflecting low torus covering factors in the \textit{Sparks} sample (section~\ref{sec:results:sed_fitting:agn-torus}).} We test the reliability of \texttt{Prospector}?s AGN torus model by comparing its predictions with optical emission-line diagnostics. We find that the model can distinguish, on average, between optically-classified AGN and purely star-forming galaxies, with AGN hosts showing a statistically significant higher contribution from the torus to their mid-infrared emission. The model infers torus luminosities that are an order of magnitude lower than expected from the AGN bolometric luminosities derived from optical emission lines, yielding a median $L_{\mathrm{torus}}/L_{\mathrm{AGN}} \sim 0.03$ compared to the typical $\sim 0.1-0.5$ found in other AGN samples. We show that the observed mid-infrared luminosities of our sources cannot accommodate $L_{\mathrm{torus}}/L_{\mathrm{AGN}}$ much larger than $0.1$, even under the extreme assumption that the AGN dominates the mid-infrared completely. This low luminosity ratio corresponds to a torus covering factor of $\sim 0.3$, smaller than the typical $\sim 0.6-0.7$ for the general AGN population, and may be a signature of their evolutionary stage where feedback has already cleared out a significant fraction of the obscuring material during the coalescence stage in a merger. 

Our results point to two key avenues for model development within \texttt{Prospector} that would be particularly beneficial for characterizing the \textit{Sparks} galaxies. First, a robust joint spectro-photometric fit of the \textit{Sparks} galaxies requires new routines to account for the physical color gradients between the fiber-sampled spectrum and the integrated photometry, using some simplified geometrical model of the galaxy. Second, disentangling AGN and starburst heating sources may benefit from including AGN emission in ultraviolet-optical wavelengths, and enforcing energy balance between the absorbed photons and those emitted by the dusty torus. At the moment, inclusion of an AGN torus component breaks the energy-balance nature of the code, as this component is not tied to an optical counterpart.

\section*{Data Availability}

We provide a catalog of far-ultraviolet to far-infrared photometry for the \textit{Sparks} galaxies online, \texttt{Sparks\_photometry}, after all the corrections described in Section~\ref{sec:SEDs:photometry}.

We provide four catalogs of the best-fitting parameters obtained from fitting \texttt{MAGPHYS} and \texttt{Prospector} to the multi-wavelength SEDs, as well as \texttt{Prospector} to the optical spectra. We further provide the best-fitting parameters for the \texttt{Prospector} fit to the multi-wavelength SEDs with the AGN torus component turned off. 

\acknowledgments{
We thank G.~Canalizo, D.~French, H.~Netzer, C.~Ricci, K.~Rowlands, and V.~Wild for providing valuable feedback that improved the presentation and interpretation of the results. 

We thank the referee for their careful reading and constructive feedback, and for their thoughtful consideration of both papers.

During the data collection and initial analysis, D.~Baron was supported by the Carnegie--Princeton Fellowship.
}

\software{Astropy \citep{astropy13, astropy18, astropy22},
		  IPython \citep{perez07},
          scikit-learn \citep{pedregosa11}, 
          SciPy \citep{scipy01},
		  matplotlib \citep{hunter07}}\vspace{2cm}

\bibliography{ref.bib}

\appendix

\onecolumngrid

\section{\texttt{Prospector} spectroscopy versus photometry fits}\label{app:sed_extra:pros-spec-photo}
\setcounter{figure}{0}

\begin{figure*}
	\centering
\includegraphics[width=1\textwidth]{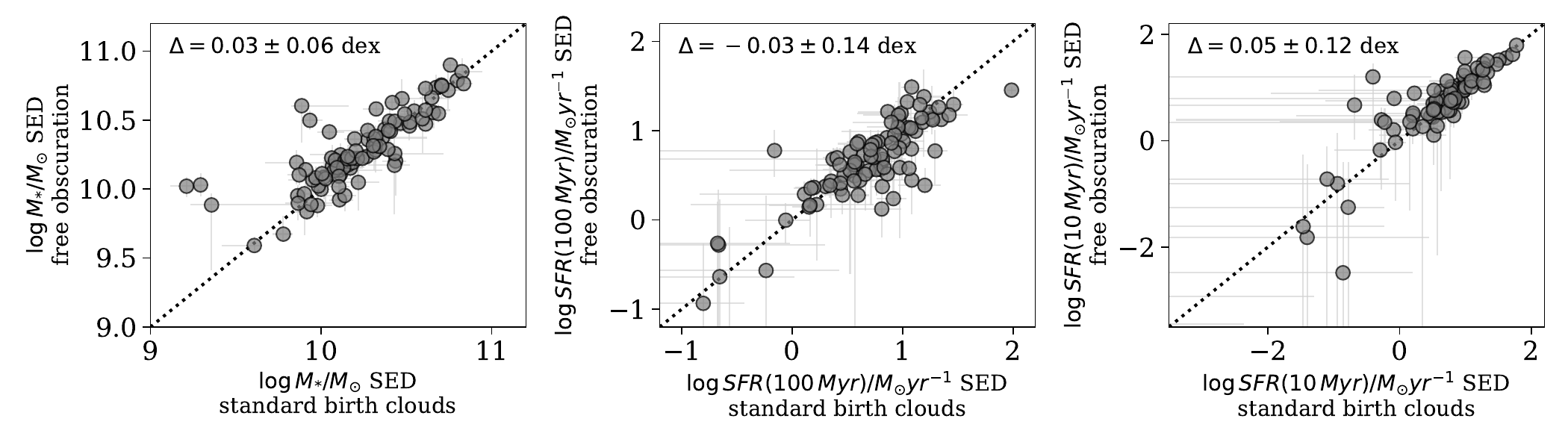}
\caption{\textbf{Comparison of best-fitting stellar masses and SFRs for the photometric \texttt{Prospector} fits under varying dust attenuation assumptions.} The panels show the stellar mass, the average SFR over the last 100 Myrs, and over the last 10 Myrs. They compare the properties derived under the free obscuration versus standard birth-clouds models. The black dashed lines mark the 1:1 relation. In each panel, $\Delta$ denotes the median and median absolute deviation (MAD) of the logarithmic difference between the properties derived into the two fits.}
\label{f:Pros_SED_comparison_fixed_vs_free}
\end{figure*}

Figure \ref{f:Pros_SED_comparison_fixed_vs_free} compares the best-fitting stellar masses and SFRs over the past 100 and 10 Myrs for the multi-wavelength SED \texttt{Prospector} fits under the `standard birth-clouds' and `free obscuration' models. The observed differences between derived properties are smaller than those we find between the spectroscopy and photometry derived \texttt{Prospector} models using the standard birth-clouds assumption. 

\begin{figure*}
	\centering
\includegraphics[width=1\textwidth]{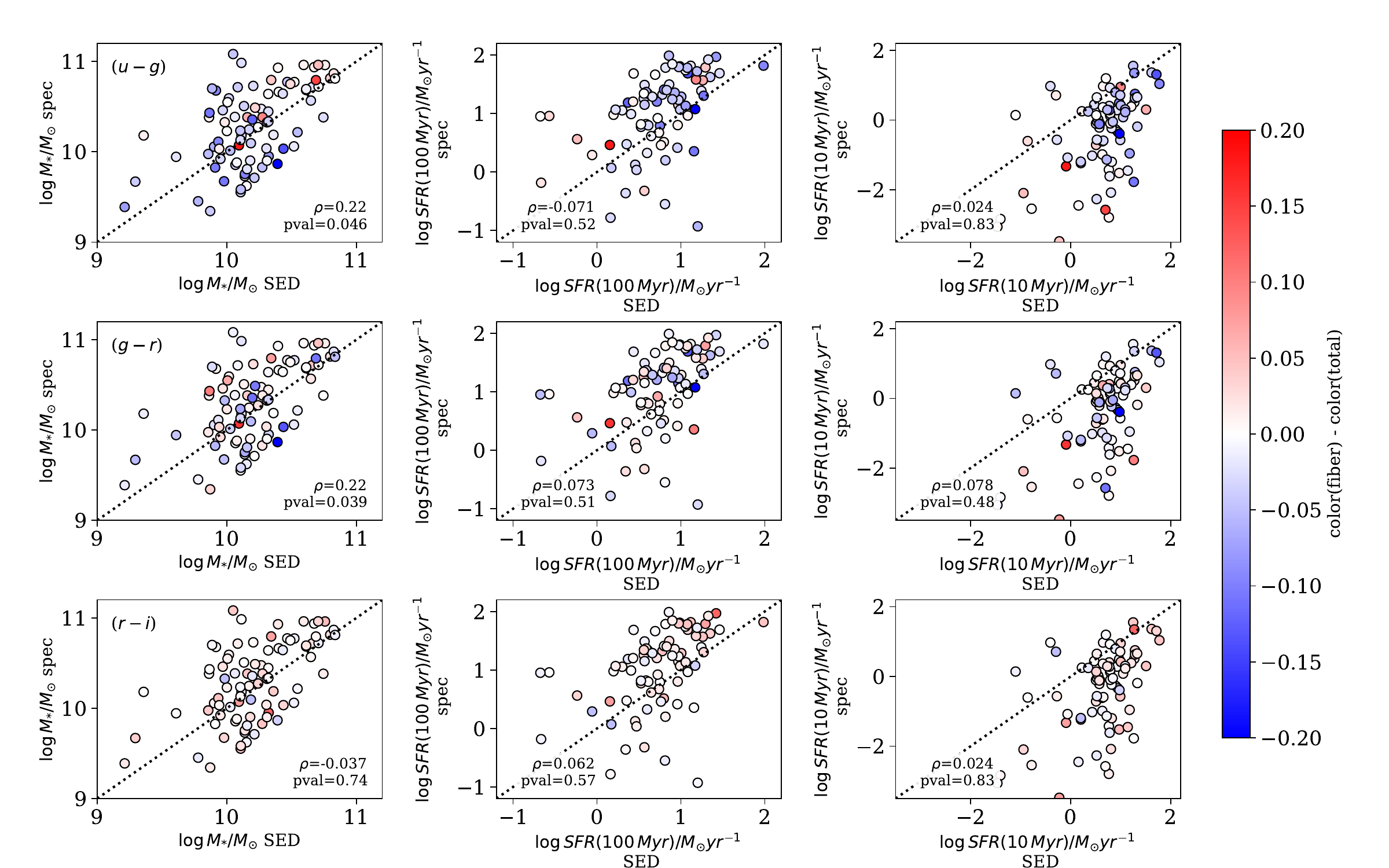}
\caption{\textbf{Comparison of best-fitting stellar masses and SFRs for the spectroscopic and photometric \texttt{Prospector} fits, color-coded by broad-band color gradients.} In each row, the panels show the stellar mass, the average SFR over the last 100 Myrs, and over the last 10 Myrs. The points are color-coded by the difference in colors observed within the fiber and outside, e.g., $(u-g)_{\mathrm{fiber}} - (u-g)_{\mathrm{total}}$ for the first row. In each panel, $\rho$ and $pval$ denote the Pearson correlation coefficient and its p-value between the logarithmic difference of the property, $\Delta y = \log(x_{\mathrm{spec}}) - \log(x_{\mathrm{SED}})$ (where $x$ is the stellar mass or SFR), and the color difference. No significant correlations are found.}
\label{f:Pros_SED_vs_spec_comparison_band_colors}
\end{figure*}

\begin{figure}
	\centering
\includegraphics[width=\columnwidth]{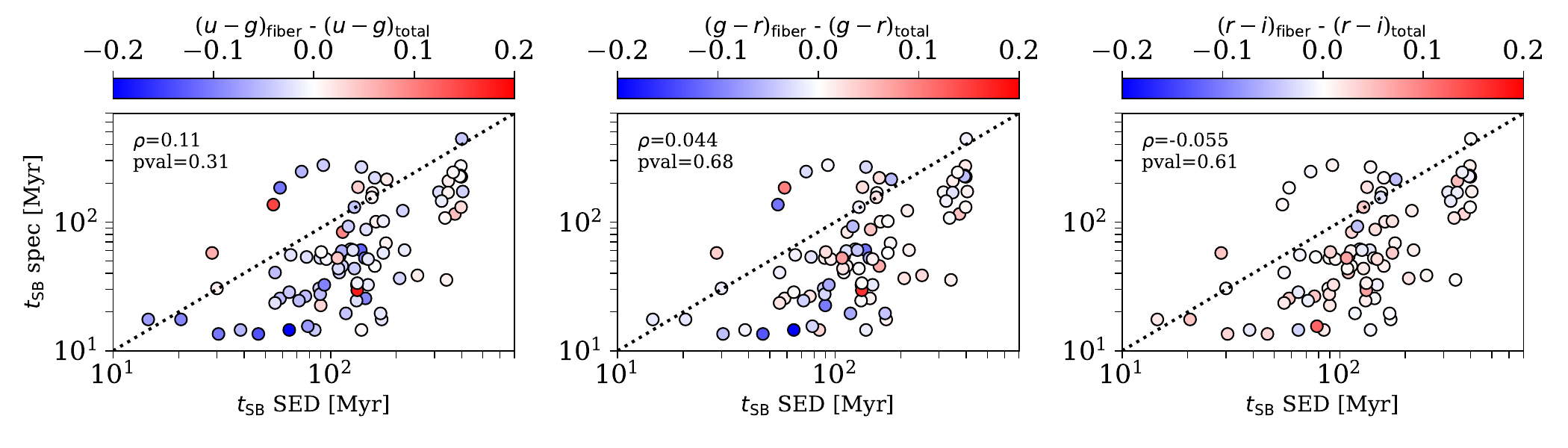}
\caption{\textbf{Comparison of starburst ages from spectroscopic and photometric \texttt{Prospector} fits, color-coded by broad-band color gradients.} The points show ages derived from the best-fitting non-parametric SFHs using optical spectroscopy versus multi-wavelength photometry. They are color-coded by the difference in colors observed within the fiber and outside, e.g., $(u-g)_{\mathrm{fiber}} - (u-g)_{\mathrm{total}}$ for the left panel. In each panel, $\rho$ and $pval$ denote the Pearson correlation coefficient and its p-value between $\Delta y = \log(t_{\mathrm{SB, \, spec}}) - \log(t_{\mathrm{SB, \,SED}})$ and the color difference. We do not find strong and significant correlations.}
\label{f:PSB_Prospector_comparison_colors}
\end{figure}

Figure \ref{f:Pros_SED_vs_spec_comparison_band_colors} compares the stellar mass and SFRs over the past 100 and 10 Myrs for the spectroscopy and photometry \texttt{Prospector} fits, color-coded by the differences in colors $(u-g)$, $(g-r)$, and $(r-i)$, within the fiber versus the total colors. We estimate Pearson correlation coefficient and its p-value between the logarithmic difference of the property, $\Delta y = \log(x_{\mathrm{spec}}) - \log(x_{\mathrm{SED}})$ (where $x$ is the stellar mass or SFR), and the color difference. We do not find significant correlations between the differences in physical properties and any of the color gradients. In figure \ref{f:PSB_Prospector_comparison_colors} we plot the starburst age derived from spectroscopy versus photometry, color-coded by the differences in colors. We do not find strong and significant correlations between these properties.

\clearpage

\section{SED fitting quality in infrared wavelengths}\label{app:sed_extra:sed_ir}
\setcounter{figure}{0}

\begin{figure*}
	\centering
\includegraphics[width=1\textwidth]{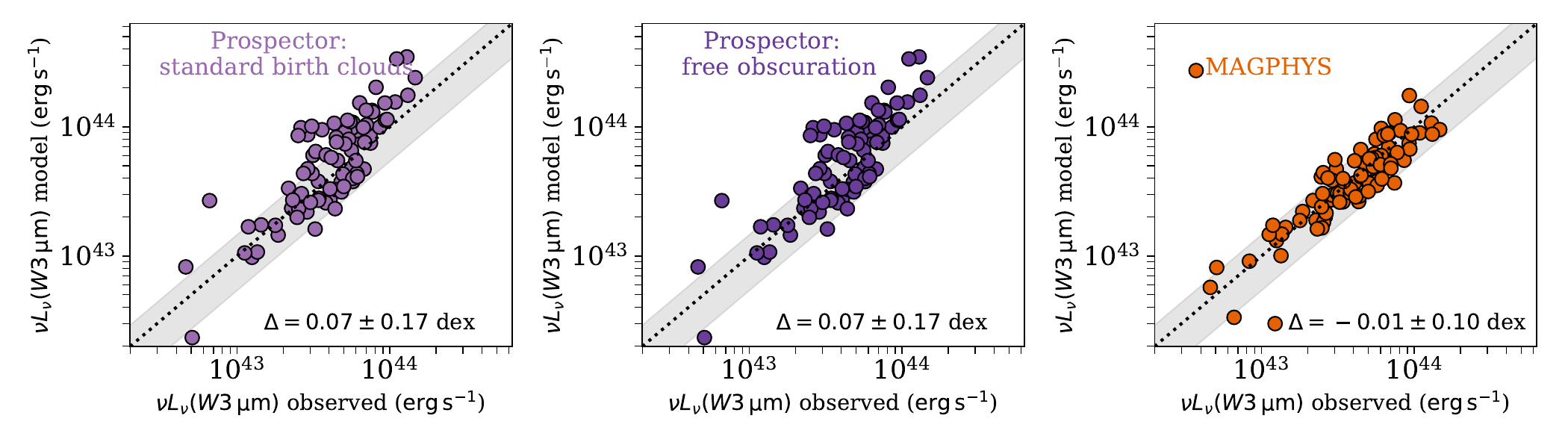}
\caption{\textbf{Comparison of predicted versus observed L(W3) for the SED fitting models.} Predicted luminosity in the WISE W3 band obtained from the best-fitting multi-wavelength SED by \texttt{Prospector} and \texttt{MAGPHYS} versus the observed luminosity. The black dashed line represents the 1:1 relation, and the light-gray band represents $\pm 0.3$ dex around it. In each panel, $\Delta$ denotes the median and MAD of the logarithmic difference between the plotted properties. }
\label{f:LW3_comparison_with_SED_codes}
\end{figure*}

\begin{figure*}
	\centering
\includegraphics[width=1\textwidth]{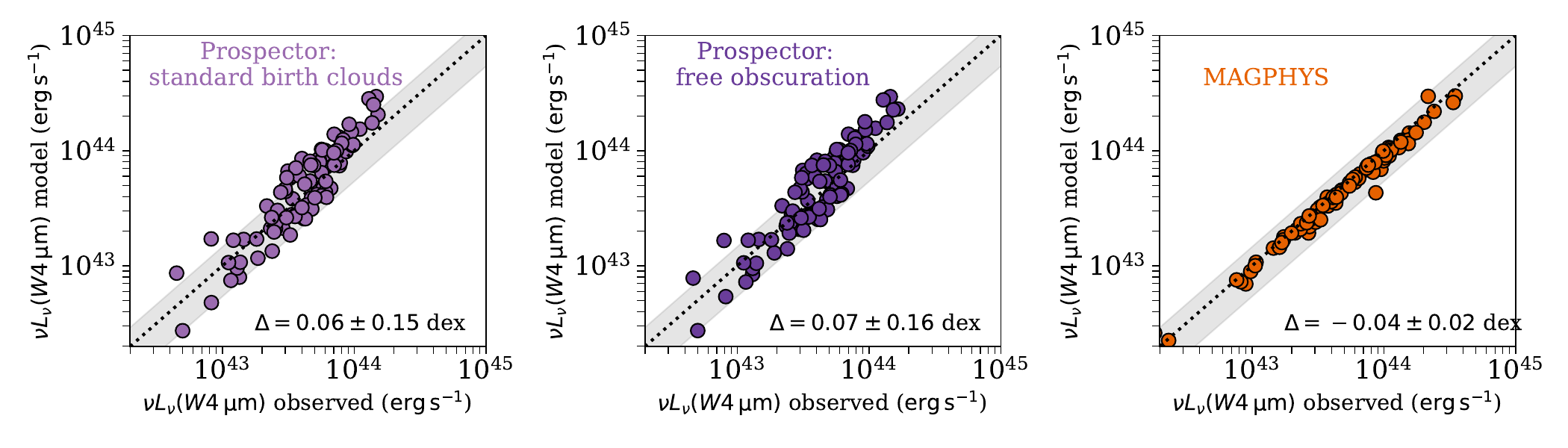}
\caption{\textbf{Comparison of predicted versus observed L(W4) for the SED fitting models.} Predicted luminosity in the WISE W4 band obtained from the best-fitting multi-wavelength SED by \texttt{Prospector} and \texttt{MAGPHYS} versus the observed luminosity. The black dashed line represents the 1:1 relation, and the light-gray band represents $\pm 0.3$ dex around it. In each panel, $\Delta$ denotes the median and MAD of the logarithmic difference between the plotted properties. }
\label{f:LW4_comparison_with_SED_codes}
\end{figure*}

\begin{figure*}
	\centering
\includegraphics[width=1\textwidth]{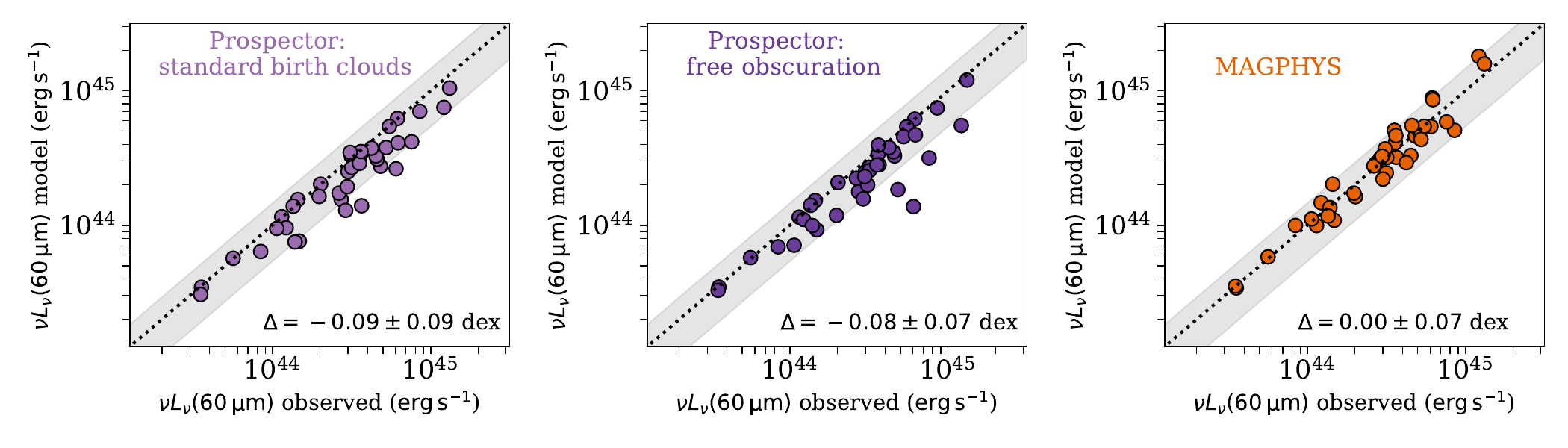}
\caption{\textbf{Comparison of predicted versus observed L(60 \mic) for the SED fitting models.} Predicted luminosity in the IRAS 60 \mic\ band obtained from the best-fitting multi-wavelength SED by \texttt{Prospector} and \texttt{MAGPHYS} versus the observed luminosity. The black dashed line represents the 1:1 relation, and the light-gray band represents $\pm 0.3$ dex around it. In each panel, $\Delta$ denotes the median and MAD of the logarithmic difference between the plotted properties. }
\label{f:L60_comparison_with_SED_codes}
\end{figure*}

Figures \ref{f:LW3_comparison_with_SED_codes}, \ref{f:LW4_comparison_with_SED_codes}, and \ref{f:L60_comparison_with_SED_codes} compare the predicted luminosities at WISE W3, W4, and IRAS 60 \mic by the SPS SED fitting code with the observed ones. 

\section{SFR(100 Myr) and SFR(10 Myr) derived from SPS fitting}\label{app:sed_extra:sfr100_vs_sfr10}
\setcounter{figure}{0}

\begin{figure*}
	\centering
\includegraphics[width=1\textwidth]{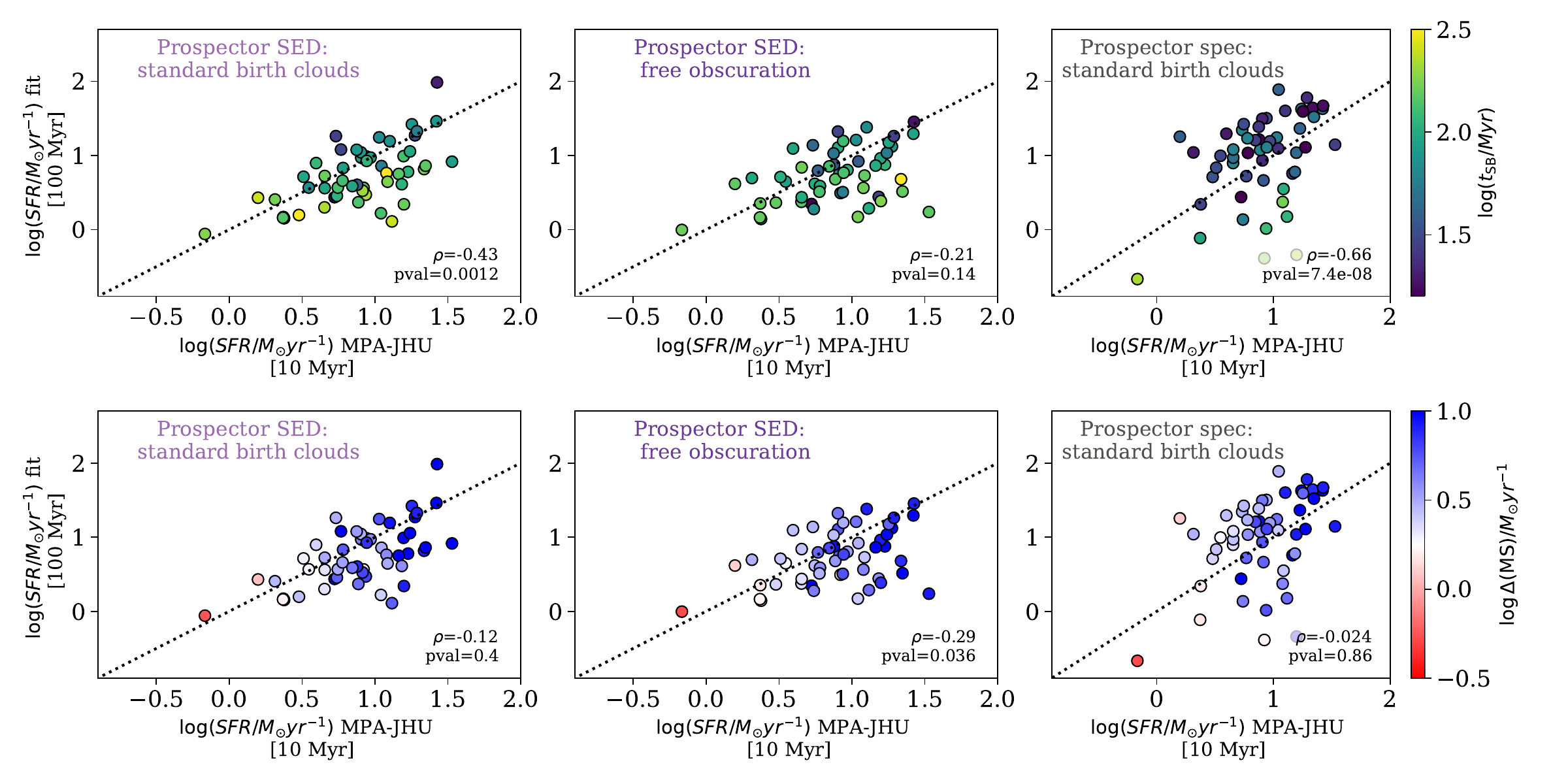}
\caption{\textbf{SFR(100 Myr) versus SFR(10 Myr), color-coded by transition indicators.} The panels compare the SPS-derived SFR(100 Myr) from the two \texttt{Prospector}+SED models to the H$\alpha$-based SDSS SFR(10 Myr) for star-forming galaxies. Points are color-coded by starburst age from the photometric fits (top row) and by offset from the star-forming main sequence (bottom row). In each panel, $\rho$ and $pval$ indicate the Pearson correlation coefficient and its p-value between $\Delta y = \log(\mathrm{SFR}_{100\, \mathrm{Myr}}) - \log(\mathrm{SFR}_{10\, \mathrm{Myr}})$ and $t_{\mathrm{SB}}$ or $\Delta(\mathrm{MS})$, respectively.}
\label{f:Pros_comparison_SFRs_different_scales_vs_SFH}
\end{figure*}

In figure~\ref{f:Pros_comparison_SFRs_different_scales_vs_SFH}, we color-code the relation between SFR(100 Myr) and SFR(10 Myr) by indicators of the galaxy?s stage in the transition from starburst to post-starburst. The top row shows color-coding by the starburst age derived from \texttt{Prospector}+SED, and the bottom by the offset from the star-forming main sequence using \citet{whitaker12}. For the \texttt{Prospector} standard birth-clouds model, we find a significant negative correlation between $\Delta y = \log(\mathrm{SFR}_{100\,\mathrm{Myr}}) - \log(\mathrm{SFR}_{10\,\mathrm{Myr}})$ and $t_{\mathrm{SB}}$, such that SFR(100 Myr) tends to fall below SFR(10 Myr) at older starburst ages.

\begin{figure*}
	\centering
\includegraphics[width=1\textwidth]{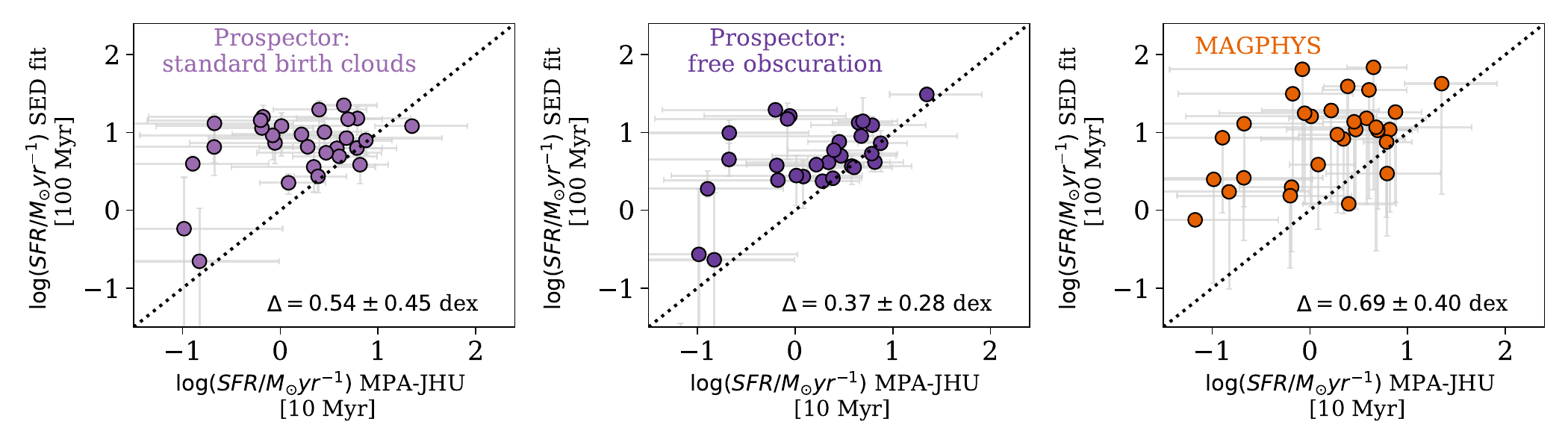}
\caption{\textbf{Comparison of SFR(100 Myr) derived using different SPS fitting codes and modeling assumptions for \textit{non} star-forming galaxies.} Each panel shows the best-fitting SFR, averaged over the past 100 Myr, from SPS fitting versus the SDSS catalog SFR. \texttt{Prospector}+spec results are not shown as we believe these do not trace the ongoing star formation accurately. Only galaxies with line ratios inconsistent with star formation are included (composites, AGN, weak lines); the SDSS SFR is based on the D$_{n}$4000\AA, index and is highly uncertain. The black dashed line indicates the 1:1 relation, and $\Delta$ gives the median and MAD of the logarithmic differences.}
\label{f:comparison_SFR_SED_fits_composites_and_AGN}
\end{figure*}

Figure \ref{f:comparison_SFR_SED_fits_composites_and_AGN} shows SFR(100 Myr) derived using SPS fitting of the multi-wavelength SEDs for the galaxies with line ratios inconsistent with being dominated by star formation ionization (composites, AGN, and weak lines). These are compared to the SDSS SFRs, which are based on the D$_{n}$4000\AA\, index, and are highly uncertain. Large offsets and scatters are seen across the different codes and assumptions. 

\end{document}